\documentclass[11pt]{article}
\usepackage{graphicx} 
\usepackage{physics}

\usepackage{amsmath}

  \usepackage[parsep]{collref}

\DeclareFontFamily{OT1}{rsfs10}{}
\DeclareFontShape{OT1}{rsfs10}{m}{n}{ <-> rsfs10 }{}
\DeclareMathAlphabet{\mathscript}{OT1}{rsfs10}{m}{n}

\numberwithin{equation}{section}
\usepackage[T1]{fontenc}
\usepackage[dvipsnames]{xcolor}
\usepackage{tikz} 
\usetikzlibrary{arrows,decorations.pathreplacing,decorations.markings,snakes}
\usetikzlibrary{cd}

\newcommand{\w}{\wedge}

\newcommand{\RR}{{\mathbf{\rR}}}

\newcommand{\al}[1]{\begin{align}{#1}\end{align}}

\def\a{\alpha}

\def\g{\gamma}

\def\z{\psi}
\def\k{\kappa}
\def\l{\lambda}

\def\n{\nu}

\def\p{\pi}

\def\r{\rho}
\def\s{\sigma}

\def\x{\xi}
\def\z{\zeta}
\def\v{\varphi}

\def\D{\Delta}
\def\G{\Gamma}

\def\L{\Lambda}
\def\O{\Omega}

\def\Q{\C}

\def\X{\Xi}

\def\bb{{\bar b}}

\def\g {\gamma}

\def\gsim{ \lower .75ex \hbox{$\sim$} \llap{\raise .27ex \hbox{$>$}} }
\def\lsim{ \lower .75ex \hbox{$\sim$} \llap{\raise .27ex \hbox{$<$}} }
\def\be{\begin{equation}}
\def\ee{\end{equation}}
\def\bea{\begin{eqnarray}}
\def\eea{\end{eqnarray}}

\def \td {\tilde}

\def \ha {{1 \ov 2}}

\def \sql {{\sqrt{\l}}\ }
\def \del{\partial}
\def \a {\alpha}
\def \aa {{\a'}}
\def\ov{\over}
\def \ci {\cite}

\def \foot {\footnote}

\def\la{\label}
\def\foot{\footnote}
\newcommand{\rf}[1]{(\ref{#1})}

\def \OO {{\cal  O}}\def \no {\nonumber}

\def \uu {{\rm u}}
\def \z {\zeta}
\def \N {{\cal N}}
\def \xx {{\rm x}}
\def \adss {$AdS_5 \times S^5$\ }

\def \C  {{\cal C}}  
\def \FF {{\cal F}}
\def \rF {{\rm F}}
\def \g {{\gamma} } 
\def \ov {\over}

\def \n {\nu}
\def \N {{\cal N}}

\def \iffa {\iffalse}

\def \iffa  {\iffalse}
\def \RR {{\mathbb R}}
\def \R {{\rm R}}
\def \s {\sigma}  
\def \G {\Gamma} 
\def \CP  {{\rm CP}}\def \gs {g_s}

\def \te {\textstyle}

 \def \rR  {{\rm R}}

\def \no {\nonumber}

\def \ss {\tau}

\def \ba {\begin{align}}
\def \ea {\end{align}}
\def \RR  {{\mathbb R}}
\def \rr {{\rm R}}

\def \T  {{\rm T}}

\def \ln {{\rm log\,}}

\def \ed {
\small
\bibliography{biblio2.bib}
\bibliographystyle{JHEP-v2.9}
\end{document}
}
\def \edo {\end{document}}

\def \vp  {\varphi }  \def \rr {{\rm r}} 
    
\def \ads {\mathrm{AdS}}
\makeatletter
\renewcommand*{\@fnsymbol}[1]{\textit{\@alph{#1}}}
\makeatother
\usepackage{amsmath,amssymb,lmodern}    
\usepackage{caption}
\usepackage{dsfont}
\numberwithin{equation}{section}
\usepackage{slashed}

\pdfoutput=1
\usepackage{amsthm, amssymb,latexsym,amsfonts}
\usepackage[T1]{fontenc}
\usepackage{comment}
\usepackage{graphicx}
\usepackage{caption}
\usepackage{amsmath}
\usepackage{appendix}
\usepackage{tikz} 
\usetikzlibrary{cd}
\begin{document}
 \begin{titlepage}
\title{ 
   { 
   Wilson loop in $\ads_3 \times S^3 \times T^4   $ from quantum M2 brane 
    }
  \\[0.5em] }
\author{\large    Arkady A. Tseytlin\footnote{Also at  the Institute  for Theoretical and Mathematical Physics    (ITMP)  of  MSU
      and  Lebedev
    Institute.}  \ \ \ \ \ 
    and \ \ \ \ Zihan Wang\thanks{zihan.wang18@imperial.ac.uk} 
       \\[0.9em]
\it      Abdus Salam Centre for Theoretical Physics \\[0.03em]
  \it   Imperial College London,  SW7 2AZ, UK 
}
  \maketitle
\begin{abstract}
\noindent
Type IIB string theory on  AdS$_3 \times S^3\times T^4$   with RR flux 
  as the near-horizon limit of the  D1-D5   solution
is  expected to be dual  to a  (4,4) supersymmetric  2d  CFT   
parametrized   by the integers  $Q_1,Q_5$ and other moduli.   It is related  by T-duality to type IIA string theory 
 in  the  near-horizon   limit of the D2-D4   solution   which 
 admits  an uplift to the 11d    AdS$_3 \times S^3\times T^5$  background  which is the   near-horizon limit of the M2-M5  solution.
 We point out that this   relation allows one  to  use  the quantum M2-brane description 
  to probe ``non-planar'' corrections in the dual 2d CFT, in close analogy with  the ABJM theory case  (described by 
 M-theory on  AdS$_4 \times S^7/\mathbb{Z}_k$). We   consider  an   analog of  a supersymmetric  Wilson loop  (line  defect) expectation value      represented by type IIA 
 string  partition function  expanded  around    AdS$_2\subset $AdS$_3$   minimal surface.
  Its   M-theory  analog   is   the M2  brane partition function  expanded  near 
    AdS$_2\times S^1$.   We compute  the  1-loop   contribution  $Z_1$  to the  M2 brane partition function 
   and find that  in contrast to the ABJM case   in  arXiv:2303.15207  (where   
 $Z_1= (2\sin{2\pi\over  k})^{-1} = {k\over 4 \pi} + {\pi\over 6k} +...$   contains  an   infinite series  of  higher  genera  string  
 corrections, $k^{-1} \sim {g_s\over \sqrt {\rm T}}$),  here   it    is given  solely    by the leading  string-theory contribution $Z_1= {\kappa \over \sqrt{2\pi}}$ where $\kappa \sim \sqrt{Q_5}$   plays the  role   analogous to   $k$.  We also   discuss  a      generalization to the  mixed flux  case  which  is straightforward    from the 11d perspective. 
 \end{abstract}

%
\end{titlepage}

\def \CP  {{\rm CP}}
\def \edo {\newpage
\small
\bibliographystyle{JHEP-v2.9}
\bibliography{bib}
\end{document}}
\def\la{\label}\def\foot{\footnote} 
\def \tG {\Gamma} 
\def \ss {{\rm s}}  \def \ff {{\rm f}} \def \gv {{\rm g}} 
\def \ddd {d} \def \N  {{\cal N}}  
\def \fin {{\rm fin}}
\def \sql {\sqrt \lambda}
\def \str {{\rm s}}  \def \W  {{\cal W}}
\def \Z  {{\cal Z}}
\def \adn {{\rm AdS}_2}
\def \adt  {${\rm AdS}_2\ $}
\def \tot {{\rm tot}}
\def \p {\phi}
\def \xx {{\rm x}}
\def \rR   {{\rm R}}
\def  \WW   {{\rm W}}
\def \R  {R^{(2)}}
\def \ym  {{_{\rm  SYM}}}
\def \rG  {{\rm G}}\def \D  {{\rm D}}
\def \OO  {{\cal O}} \def \bG  {\bar \G}
\def \na {\nabla}\def \rr {{\rm r}}\def \cc  {{\rm c}}
\def \lm {v}
\def \CP  {{\rm CP}}
\def \g {{\rm g}_{\rm s}} \def \T {{\rm T}}

 \def \adss {AdS$_5 \times S^5$} \def \adsc {AdS$_4 \times \CP^3$}  \def \adst {AdS$_3 \times S^3\times T^4$}
  \def \adsp {AdS$_3 \times S^3$}  \def \adsss  {AdS$_4 \times S^7/\mathbb{Z}_k$}
  \def \lpl {\ell_{_{\rm P}}}
  \def \RR {R^{(2)}} 
  \def \O {{\cal O}}
\def \n {{\rm p}}
\def \C {{\cal Q}}\def \v {{ v}}
  \def \ls {\ell_s}  \def \gs {g_s} 
\def \AdS {{\rm AdS}}

 \def \adsts {AdS$_3 \times S^3\times T^5$}
\def \y {{\rm y}}  \def \w {{\rm w}}
\def \vol {{\rm vol}}
\def \x {\vp} \def \AdS {{\rm AdS}}
  \def \s {\sigma} \def \fo {\tfrac{1}{4}}
\def \ssigma {{\bf \sigma} }
\def \M  {{\cal M}}
  \def \z {{\cal Z}} \def  \ha {\tfrac{1}{2}}
\def \bGamma {\bar \Gamma} 
\def \tphi {{\tilde \phi}}
  \def \vv {{\rm v}}
  \def \uu {{\rm u}} 
  \def \cc {{\rm c}}
  
  \def \aa {{\rm a}}  \def \bb {{\rm b}}
  
   \def \k {{\kappa}}
   \def \vkappa  {\varkappa}
   \def \det {{\rm det\,}}  
\def \ka {\vkappa}  \def \hv {\hat v} 
\def \ww  {{\rm w}}

\def \ftr{1/3}
\def\ftt{{2}/{3}}
  \def \GG   {{\rm G}} 
\def \FF {{\rm F}} \def \H  {{\rm H}} 
   \def \hv {\hat v} 
\def \Q  {{\cal Q}} 
\def \qq  {{\rm q}}\def \www  {\tilde w}  \def \kkk  {\tilde \k}  \def \ga {\gamma} \def \r   {\rho}
\def \Z  {\mathbb{Z}} \def \pp {{\rm p}}  \def \H  {{\rm H}}  \def \FF {{\rm F}}  \def \bC   {{\bar \C}}
\def \tL {\tilde L}   \def \tk {\tilde \kappa}  \def \tR {\tilde R}
\def \tR {\tilde R} \def \X {{\Lambda}}
\def \tT {\tilde \T}
 \def \hq {\hat q}
\def \z  {\zeta} 
\def \rB  {{\rm B}}   \def \rF {{\rm F}}

\def \l  {\lambda}
{\small 
\tableofcontents }

  \section{Introduction} 
 
 Most  symmetric examples of AdS/CFT \cite{Maldacena:1997re,Aharony:1999ti,Aharony:2008ug}
 related to \adss, \adsc\  and \adst\  share the  common feature that their spectrum 
 is expected to be controlled by integrability in the large $N$ (tree-level  string) limit. 
 Going beyond  this planar limit  at strong coupling  is challenging as this  requires  computing string loops. 
 
 Remarkably, in the   \adsc\  or     ABJM  theory  case  one is able to   compute certain   non-planar  strong-coupling 
corrections using  its M-theory  uplift  in terms of quantum M2 branes in \adsss\ \ci{Giombi:2023vzu,Beccaria:2023ujc,Giombi:2024itd,Beccaria:2025vdj,Beccaria:2025npl,Gautason:2025per,Gautason:2025plx,
Gautason:2025bft}.
 Here we point out that an   analogous   approach is possible  also   in the   \adst\  type IIA case that  
 admits  an  11d M-theory embedding. 
 
 The standard   discussion    of the AdS$_3$/CFT$_2$   duality\foot{We will be mostly  interested  in the  case of 
 \adst\  supported  by RR flux.}
   starts with the D1-D5   combination in type IIB   string theory 
  that has \adst\ as its near-horizon  limit.  Guided by 
   weak-coupling (flat-space)  open-string description of D1-D5  combination 
  one expects that the dual 2d CFT 
    should be related to an IR limit of 
  a  coupled system of  three  (4,4) supersymmetric  multiplets 
  (see, e.g.,  \ci{Witten:1997yu,Aharony:1999ti,OhlssonSax:2014jtq}): $U(Q_1)$  adjoint vector multiplet, $U(Q_1)$  adjoint  hypermultiplet, 
and  $U(Q_1)\times U(Q_5) $ bi-fundamental  hypermultiplet. It should 
   correspond to a certain deformation of symmetric 
  orbifold  $(T^4)^{N}/S_N$   2d CFT  where $N= Q_1 Q_5  $  (see, e.g.,  \ci{Seiberg:1999xz,David:2002wn,Aharony:2024fid}). 

In general,  details  of this duality (in particular, interpolation from weak to strong coupling) 
 beyond the BPS-protected sector  are  still poorly understood 
 (for some  progress see, e.g.,    \ci{Babichenko:2009dk,Demulder:2023bux,Frolov:2025tda,Ekhammar:2026ykk}  and  also \ci{Gaberdiel:2024dfw,Chester:2024wnb,Jiang:2025oar,Jiang:2026xnh}). 

The  simplest  ``protected'' observable  is the  free energy on $S^2$  related to the central charge. 
The central charge count on the  gauge theory side\foot{The contribution to 2d conformal 
anomaly of  a single free  (4,4)  hypermultiplet (4 real scalars and 4 real fermions) is 
$\cc=6$. 
 The  2d vector multiplet has an  irrelevant kinetic term and thus contributes  to anomaly  only through 
ghost  factor, with a single  vector    giving    $\cc=-1$.
The $U(1)$  part of the  vector multiplet   is decoupled (representing the c.o.m. of the bound D1-D5 combination), so that 
finally \   $\cc =6 Q_1 Q_5  +  Q_1^2    - 6 (Q_1^2-1) = 6 Q_1 Q_5 + 6 $. 
The same result is found  by counting  the $SU(2)$ chiral anomaly  
of  the (4,4)  superconformal algebra \ci{Witten:1997yu,Aharony:1999ti}.}
 gives $\cc =6 Q_1 Q_5 + 6 $. 
On  the  string theory  side the  free energy   is proportional to the   logarithmically IR divergent volume of AdS$_3$  and  is thus  related 
to the  UV conformal anomaly   in the boundary theory. 
 The leading $6Q_1Q_5$ term   is  reproduced  by  the classical action of  the  10d 
  supergravity on AdS$_3 \times   S^3 \times T^4$  \ci{Henningson:1998gx}  while 
   the subleading  $+6$ term   comes  from   the 1-loop  10d supergravity  contribution  \ci{Beccaria:2014qea}
   (with all subleading in large $N=Q_1 Q_5 $  contributions expected to cancel out).\foot{This is  analogous 
   to the $N^2\to N^2-1$  shift in \adss  case \ci{Beccaria:2014xda}   and  the $4 N^3 - \frac{9}{4}  N  \to   4 N^3 -  \frac{9}{4} N - \frac{7}{4} $
   shift in the AdS$_7 \times S^4$  case  \ci{Beccaria:2014qea}.}
 
 Another   special   observable that will be our main  focus here is the expectation value of 
  a supersymmetric conformal 1-dimensional defect operator  which is 
 a formal    analog of the $\ha$-BPS Wilson loop  (WL)  in  gauge theories and     on the 
    string theory  side  should   be represented  by  the  partition function  expanded near the 
 AdS$_2\subset \AdS_3$  minimal surface  \ci{Drukker:2000ep,Giombi:2020mhz,Bliard:2024bcz}.

Performing T-duality along one   D5 direction  (a circle in $T^4$)  one   gets the 
D2-D4 combination in type IIA   string theory  that also has 
\adst\  space (supported by the RR  4-form   flux) as its near-horizon limit. 
The corresponding dual 2d   CFT should be   
 equivalent  to the one in the type IIB   case 
up to a  redefinition  of parameters (in particular, $Q_5\to Q_4$, $Q_1\to Q_2$). 

The key point   here   will be  that   the 
  D2-D4 solution  admits  the uplift to the M2-M5    solution of 11d supergravity \ci{Tseytlin:1996bh} 
 that   has the AdS$_3 \times S^3 \times T^5$ as its  near-horizon limit. 
 This   opens up a possibility to use
  the  quantum M2 brane probe  method   to shed  light on non-planar corrections in 
 the dual 2d  CFT. 
 
  In particular, 
 the      AdS$_2$  minimal surface  representing WL in type IIA   
 string on AdS$_3 \times S^3 \times T^4 $   supported by  RR   flux 
 gets   uplifted to  AdS$_2\times S^1 $    M2  brane  solution  in  AdS$_3 \times S^3 \times T^4 \times S^1$
 where M2  is wrapped on  $S^1\subset T^4$. 
 This  is     analogous to  how  $\ha$-BPS  WL in ABJM theory  is  represented by the 
  AdS$_2\times S^1 $   M2  brane in \adsss\ \ci{Sakaguchi:2010dg,Giombi:2023vzu}.
  
 Our aim  below will be to compute the 1-loop M2  brane  correction  to the expectation value of the  circular WL 
 in AdS$_3 \times S^3 \times T^4 \times S^1$ 
  following the same  steps  as in the \adsss\ case  in  \ci{Giombi:2023vzu}.


\subsection{Review}

Let us first review  the known     expression for the  1-loop WL  expectation value at strong coupling 
  represented by the string 
 partition function on the disc  that can be computed    using semiclassical expansion near the AdS$_2$ minimal surface  
\ci{Giombi:2020mhz}.  
The  three cases  of a string  with  ${\rm AdS}_2$ induced geometry in  ${\rm AdS}_3$ subspace of ${\rm AdS}_\n \times M^{10-\n}$  supported by RR fluxes, i.e.   
 \ 
 
(i)  $\n=5$: \ \   ${\rm AdS}_5 \times S^5$;  \ \ \ \ \ \   (ii) $\n=4$: \ \  ${\rm AdS}_4 \times   \CP^3$; \ \ \ \ \ \ 
(iii) $\n=3$: \ \  ${\rm AdS}_3 \times   S^3\times T^4$ \ 

\noindent
can be treated on an equal footing.
The main parameters  are  the AdS$_\n$  scale $L$, the string coupling $\g $   and the effective dimensionless 
string tension $\T$:
\be\la{1}   \T =L^2 T =  {L^2\ov 2\pi \a'}  \ .  \ee
The WL expectation   value    is   given 
 by the  string partition function  with a disk-like world sheet ending on a circle at the boundary of AdS$_\n$
\begin{align}\la{2}
 \langle \W\rangle   
 &= {1\ov \g  }
   {\rm Z}_1  +  \OO ( \g ) = W_1 \,  \big[1 + \OO( \T^{-1})\big] +   \OO ( \g )   
   \, , \\
       { \rm Z}_1 
   &=  \int [dx...]  \ e^{ -  T \int d^2 \s\,   {\cal L} }\,= e^{-\tG}, \qquad  \ \ \tG=\tG_0 + \tG_1 +  \tG_2 + ...\ . \la{20}
\end{align}
Here in \rf{2}  we separated the 
 $1/\g$ factor  corresponding  to   the leading  disk topology  contribution. 
$\tG_0= - 2 \pi \T $  is the  classical string action
 proportional to  the renormalized  ${\rm AdS}_2$  volume $\vol({\rm {\rm AdS}_2})= - 2 \pi$.  
 The 1-loop   correction 
 ${\tG_1}= \OO(\T^0)$   found from the expansion of the  Green-Schwarz  string   action  in a static and $\vkappa$-symmetry  gauge  is   given  by  the  determinants  of the fluctuation operators in AdS$_2$  world-sheet geometry 
 ($\RR= -2$) \ci{Drukker:2000ep,Giombi:2020mhz}
\begin{align}
e^{-\tG_{1}} &
=  \frac{\big[{\rm det} (-\nabla^2 +\frac{1}{4}\RR +1) \big]^{ \n-1    }  \ \big[{\rm det} (-\nabla^2 +\frac{1}{4}\RR) \big]^{5-\n} }{[{\rm det} (-\nabla^2 +2  )]^{{1\ov 2 } \n - {3\ov 2}}  \    [{\rm det} (-\nabla^2 + \RR + 4  ) ]^{1\ov 2} 
\      [{\rm det} (-\nabla^2 )]^{5 -{1\ov 2} \n}}
\no \\
&=\frac{\big[{\rm det} (-\nabla^2 +\frac{1}{2}) \big]^{ \n-1    }  \ \big[{\rm det} (-\nabla^2 -\frac{1}{2}) \big]^{5-\n} }{\big[{\rm det} (-\nabla^2 +2  )\big]^{{1\ov 2} \n-1}  \ \big[{\rm det} (-\nabla^2 )\big]^{5-{1\ov 2} \n}}\ . 
\la{3}
\end{align}
 In the  case of ${\rm AdS}_3 \times   S^3\times T^4$     supported by RR flux 
there are  4  massless  fermion  modes (which are partners of  $T^4$ bosonic modes)
  and 4 massive ones \ci{Drukker:2000ep} (the result is the same in  T-dual  type IIB  and  type IIA   embeddings). 
For general $\n$  one finds that 
\begin{align}  \la{4}
&\Gamma_1 =\ha  \sum \log \det \Delta= \G_{1\infty}   + \bG_1  \ , 
 \qquad \ \ \  \G_{1\infty} =  -   \zeta_{\rm tot}(0) \log  \Lambda  \ , \qquad    \zeta_{\rm tot}(0)=1 \ , \\ &
\bG_1 =  -\ha  \zeta'_{\rm tot}(0) =   \ha  ( \n-4) \ln (2 \pi)    \ , \la{5}
 \end{align}
where $\Lambda$ is a 2d UV cutoff (the radius of AdS$_2$ is set to 1).
  As was argued in  \ci{Giombi:2020mhz}  the  universal   divergent term  $\G_{1\infty} =  -   \log  \Lambda $
\rf{4}  should be cancelled against 
the contribution of the GS string path integral measure   so that   at the end   it gets effectively  replaced by 
a finite term 
\be 
 \G_{1\infty}  \ \ \to \ \  - \log {\sqrt {\T}\ov \sqrt{2\pi}} \ . \la{44} \ee
 As a result, the 1-loop   contribution $W_1$ in \rf{2}   is  given  by
 \be 
\la{6} 
W_1 =  {1\ov \g} {\sqrt {\T}\ov \sqrt{2\pi}}  \, e^{-\bar \G_1}\,   e^{2\pi \T}= 
    c_\n  { \sqrt \T \ov \g  }\  e^{ 2\pi \T  }          \ , \qquad \ \  c_\n =  { 1 \ov ( \sqrt{ 2 \pi} )^{\n-3} } \ .  \ee
    In  the  $\n=3$ case corresponding to ${\rm AdS}_3 \times   S^3\times T^4$
  we get $c_\n=1$. 
  
The expression \rf{6}  matches  the known    gauge theory localization results in  the 
$\n=5$  (SYM) \ci{Erickson:2000af,Drukker:2000rr}  and $\n=4$  (ABJM) \ci{Drukker:2010nc} 
  cases where   one has 
  \begin{align}
&{\rm SYM}:\   \ \ \ \ \    \g  = \frac{g^2_{_{\rm YM}}}{4\pi}=\frac{\l}{4\pi N}\ , \ \qquad
     \l = g^2_{_{\rm YM}} N  , \  \qquad  \T    = {\sql \ov 2 \pi}  \ , \la{000}\\
     &    {\rm ABJM}:\  \ \ \  
\g  = {   \sqrt \pi\,(2 \l)^{5/4}\ov  N} , \qquad  \ \ \ \ 
     \l = {N \ov k}  \ , \ \qquad  \ \ \ \   \T ={ \sqrt{ \l }\ov \sqrt 2}\ ,\ \ \  \qquad k^{-1} = {\g\ov \sqrt {8\pi\, \T}} \ . \la{101}  \end{align}
         Higher genus  string corrections to \rf{2}  
  are expected  to have the structure  \ci{Giombi:2020mhz} ($\chi$ is the  2d  Euler number)
 \begin{align}   
 \langle \W\rangle =&\sum_{\chi=1,-1, ...}  {\rm b}_\chi  \,    \Big({{\sqrt \T} \ov \g}\Big)^\chi \,    e^{ 2\pi \T  } \ \Big[1 + \OO (\T^{-1})\Big]\no \\  =  &  \frac{\sqrt{\T}}{\g}  e^{ 2\pi \T  } 
\Big(
c_\n+ c_{1\n} \frac{\g^2}{\T} \big[1 +  \O(\T^{-1})\big]
+ c_{2\n}  \frac{\g^{4}}{\T^{2}}  \big[1 +  \O(\T^{-1})\big]
+... 
\Big)\   ,  \la{7} \end{align}
  which  is indeed consistent with  that   of the  $1/N$, strong coupling  expansion  found from localization 
  in the corresponding  4d and 3d gauge theories in    \ci{Drukker:2000rr} and in \cite{Drukker:2010nc}.

In the  AdS$_4$ or ABJM case one is  able also to reproduce the leading  large $\T$ corrections  to all orders in $\g$  
by  semiclassically quantizing   M2 brane  \ci{Bergshoeff:1987cm} 
 wrapped  on $\AdS_2$    and  the 11d circle  of   radius  $\sqrt{2\ov k}$   in \adsss\   \ci{Giombi:2023vzu}.
Localization predicts \ci{Klemm:2012ii}  that  the $\ha$-BPS WL  expanded  in large $N$  at fixed $k$  is given by 
 (cf. \rf{6},\rf{101})
\begin{align}  \la{33}
  \langle  \W  \rangle =&\frac{1}{2\, {\sin}\, \frac{2\pi}{k}}\, e^{\pi \sqrt{\frac{2N}{k}}}\Big[
1 
+\O(\tfrac{1}{\sqrt N})
\Big] =   { 1  \ov 2   \sin \big( \sqrt\frac{\pi}{2}\,\frac{\g}{\sqrt \T}\big)   } \, e^{2\pi \T}  \Big[1+ \O(\T^{-1})\Big] \\
= & 
 \frac{1}{\sqrt{2\pi}} \frac{\sqrt{\T}}{\g}  \, e^{2\pi \T} \, 
\Big(
1+\frac{\pi}{12}\frac{\g^2}{\T} \big[1 +  \O(\T^{-1})\big]
+\frac{7\pi^{2}}{1440}\frac{\g^{4}}{\T^{2}}  \big[1 +  \O(\T^{-1})\big]
+O(\g^{6})
\Big) \ . \la{4444}
\end{align}
Written in terms of the AdS$_4$ radius $L$  and the radius  $R= 2 L$ of $S^7$,  the effective   M2  brane tension  and its 
 classical  action  on AdS$_2\times S^1$  is the same as the   string action  appearing in the exponent in \rf{33}
\begin{align} 
\T_2 =   L^2 R\, & T_2= {1\ov \sqrt{8} \pi} \sqrt{N k} 
\ , \qquad \qquad
L^6 = \ha \pi^2 N k \lpl^6, \ \ \qquad  T_2= {1\ov (2\pi)^2 \lpl^3}\ , \la{8} \\
& S_{_{\rm M2}}=( - 2\pi )\, ({ 2\pi \ov k})\,  \T_2 = - \pi \sqrt{ 2 N\ov k}\ . \la{88}
\end{align}
  The M2  brane semiclassical  expansion    is done  for   large $\T_2$ for fixed  $k$ (fixed 11d  background), i.e. the 
  $\frac{1}{2\, {\sin}\, \frac{2\pi}{k}}$ prefactor  in \rf{33}  should be the 1-loop M2  brane contribution  
  (that  according to \rf{7} and  \rf{44} sums up all  the leading  large   string tension   contributions at each order in $\g$). 
  The 
  $\O(\frac{1}{\sqrt N})$ term in \rf{33} 
   should correspond to the 2-loop  contribution. 
   
  Indeed, summing up  the contributions of the M2  brane fluctuations  near  AdS$_2\times S^1$ 
  one   finds  a  generalization  of the string 1-loop   contribution in \rf{3} \ci{Giombi:2023vzu}
  \begin{align}
 &\qquad \qquad \qquad   \langle \W\rangle = e^{-\hat \G_1}  e^{-S_{_{\rm M2}}} \big[1 + \O({\T_2}^{-1})\big] \  , 
 \qquad \ \  \ \ \ \hat \G_1  =- \ha \sum_{n=-\infty}^\infty \log  X_n \ , \la{9}\\
&  X_n  =\frac{\big[{\rm det}(-\nabla^2+\frac{R^{(2)}}{4}+(\frac{kn}{2}+1)^2)\big]^{\frac{3}{2}}
\,
\big[{\rm det}(-\nabla^2+\frac{R^{(2)}}{4}+(\frac{kn}{2}-1)^2)\big]^{\frac{3}{2}}
\,
{\rm det}(-\nabla^2+\frac{R^{(2)}}{4}+(\frac{kn}{2})^2) 
}{{\rm det}\big(-\nabla^2+\frac{1}{4}(kn-2)(kn-4)\big)\,
\big[{\rm det}\big(-\nabla^2+\frac{1}{4}kn(kn+2)\big) \big]^3}. \la{10}  
\end{align}
Here $n$ labels  the modes of the M2 brane fluctuations expanded in $\s_3$  (identified with 
$S^1$ coordinate with radius $\sqrt{2\ov k}$) 
  that propagate on AdS$_2$. The    $n=0$
  level corresponds  to the string mode contribution in \rf{3}  in the AdS$_4$ case  ($\n=4$).
  
  In contrast to the string case in \rf{4},\rf{5}   
   $\hat \G_1$   in \rf{9}  that corresponds to a 3d theory     should  not contain log UV   divergences, i.e. 
  should    be  given   by a finite expression  (once computed  using  $\zeta$-function regularization
   that removes  power divergences). 
   Indeed, the  coefficient   of the $\log \L$ term in \rf{4}  here is replaced by 
   \begin{equation}
   \hat \G_{1 \infty} = - \zeta_{\rm tot}(0) \log \Lambda \ , \qquad \qquad 
\zeta_{\rm tot}(0) =     \sum_{n=-\infty}^{\infty} 1  = 1 +  2\zeta_R(0) = 0\,. \la{225}
\end{equation}
   The remaining    terms take the form (for $k >2$)  \ci{Giombi:2023vzu}
     \begin{align}
\hat \Gamma_1 =  \sum_{n=1}^{\infty} \log\Big(\frac{k^2n^2}{4}-1\Big)
=&2\sum_{n=1}^{\infty} \log \frac{kn}{2} + \sum_{n=1}^{\infty}\log\Big(1-\frac{4}{k^2n^2}\Big) 
=  - \log \frac{k}{4\pi}   + \log \Big(\frac{k}{2\pi}\sin \frac{2\pi}{k}  \Big) \ ,\la{011} \\ 
&    e^{-\hat \G_1} =   \frac{1}{2\sin\frac{2\pi}{k}}
\,, \la{11}
\end{align}
where we used    that  $\sin(\pi x)=\pi x \prod_{n=1}^{\infty}(1-\frac{x^2}{n^2})$  and 
\begin{equation} 
2\sum_{n=1}^{\infty}\te  \log\frac{kn}{2} = 2 \zeta_R(0)\log\frac{k}{2}
-2 \zeta'_R(0) = -\log\frac{k}{4\pi}\,, \qquad \ \  \la{12}  \zeta_R(0)= - {1\ov 2}, \ \qquad  \zeta'_R(0)= - {1\ov 2} \log (2 \pi)\ . 
\end{equation}
Eq. \rf{11}   reproduces the prefactor in \rf{33}, and, in particular, the leading string  1-loop factor  in \rf{6} (with $\n=4$)  and   \rf{44}. 

Thus the M2 brane ``uplift''  of the GS string   \ci{Duff:1987bx}  effectively  provides  its  consistent definition, i.e.
the required  measure contribution 
that cancels the $\log \L$ term  in \rf{4}  is at the same time replacing it   with  the  required 
$\sqrt{ \T\ov 2\pi}$ term in \rf{6}, in agreement with \rf{44}  (which here automatically 
 includes the ${1\ov \g}$ factor, cf. \rf{4444}). 
 
As  we shall see  below, a similar result is found also in the    AdS$_3$  case, 
where the M2   brane probe  will be wrapping AdS$_2\times S^1$   in  AdS$_3  \times S^3\times T^4 \times  S^1$.

\subsection{Results}

Our starting point will be  the ${1\ov 4}$-BPS  M2-M5   solution  of 11d supergravity \ci{Papadopoulos:1996uq,Tseytlin:1996bh} 
with the  harmonic functions 
\begin{align} 
 &   H_{2}=1+\frac{\C_{2}}{r^{2}},\qquad H_{5}=1+\frac{\C_{5}}{r^{2}}, \qquad\qquad   \C_{2}={\lpl^{2}\hv^{-1} Q_{2}},\qquad \qquad \C_{5}={\lpl^{2} u^{-1} Q_{5}} , \label{24}\\
 & \hv\equiv v_1 v_2 v_3  v_4\ , \qquad \qquad 
 y_p\equiv y_p + 2 \pi v_p\  , \qquad  w\equiv w+ 2\pi u \ , \la{244}
\end{align}
where  
$y_p$  $(p=1, ..., 4) $  are  the  angular coordinates  (with dimensionless radii $v_p$)   of 
 $T^4$ 
    along M5 not shared with M2, $u$ is  a dimensionless
  radius of $S^1$ coordinate $w$ 
  of M2  not shared  with M5 
   and $Q_2$ and $Q_5$  are integers  (see section 2).\foot{ 
   The  dimensionless   moduli parameters  $\v, \ v_4$ and   $u$  that will be related to similar parameters of 
   D2-D4 and D1-D5   10d solutions   (cf. \ci{OhlssonSax:2018hgc})
      may  be set to 1   but it is useful   to keep  their  values   generic.}
   
   The M2-M5   solution  reduces to D2-D4  solution of type IIA theory provided the 11th coordinate is identified  with  
   one of the $T^4$ coordinates (denoted  below  as $y_4$), so that $v_4$ is  related to the  asymptotic value of the type IIA 
    string coupling $\gs$ (see Appendix \ref{apb}) 
    \be  \la{204} R_{11} = v_4\lpl =  \gs \ls \ ,\qquad \ \  v_4 = \gs^{\ftt} \ , \qquad \qquad 
     \lpl= \gs^{\ftr} \ls, \qquad 
    \ls=\sqrt{\a'}\ . \ee
        In the near-horizon limit the 11d metric becomes  that  of  AdS$_3  \times S^3\times T^5$
    \begin{align}
&ds^{2}_{11}=L^{2} \big(ds_{{\rm AdS}_{3}}^{2}+ds_{S^{3}}^{2}\big) +R^{2}dy_{p} dy_p+ R^{2}_w dw^2\ ,
 \la{25} \\
&L^{2}= (\C_{2}\C_{5}^{2} )^{1/3} \ , \qquad   R^{2}=\lpl^2 (\C_{2}\C_{5}^{-1})^{1/3}  \  ,
 \qquad R^2_w =  \lpl^2  (\C_{2}\C_{5}^{-1})^{-2/3}  \ .  
\label{255}
\end{align}
The effective dimensionless  tension  and the action  of   the M2  brane 
wrapped on ${\rm AdS}_2\subset{\rm AdS}_3$    and $y_4\equiv y_4 + 2 \pi   v_4$ 
 are given by   (cf. \rf{8})  
\be 
\T_2=L^2 R \,  T_2  =\frac{1}{(2\pi)^2}  {1\ov \sqrt{ \hv u} } \sqrt{{Q_2 Q_5}} \ , \qquad \ \ \ 
S_{_{\rm M2}}=( - 2\pi )\, ({ 2\pi v_4})\,  \T_2 = -   { v_4\ov \sqrt{\hv u} } \sqrt{{Q_2 Q_5}}   \ . 
    \label{28}
\end{equation}
Let us introduce the parameter  $\k$  proportional to $v_4^{-1} $ 
 that is the  analog of  $k$   which determines 
 the   square of effective radius  ${2\ov k} $ of the  11th angular  direction  in 
the  AdS$_4 \times S^7/\mathbb{Z}_k$ case, i.e. 
\be 
 {1\ov \k_{_{\AdS_4}} }
  =  {R_{_{S^7}}\ov L_{_{\rm AdS_4}} } {1\ov k} ={ 2\ov k}\ \   \ \  \to  \ \ \ \   {1\ov \k} \equiv  {R \ov L}  v_4 = {v_4\,\lpl \ov   \sqrt{\C_5} }\ , \qquad \qquad \k= {1\ov  v_4\,  \sqrt u}     \sqrt{Q_5}
\ .
   \la{29} 
 \ee 
 The expansion  in large $\T_2$  for fixed $\k$ (and moduli $ \hv, v_4, u$) is thus the expansion in large $Q_2$ for fixed $Q_5$. Taking $\k$ large   corresponds to shrinking the size $v_4$ of the  11th dimension or taking  the type IIA 
 string coupling $\gs$ in \rf{204} small.

The  computation  of the 1-loop correction  to  the M2  brane    partition function described in sections 3 and 4 below  leads to
the following   analog of the   AdS$_4 \times S^7/\mathbb{Z}_k$  expressions in  \rf{9},\rf{10}   
case   
\begin{align} 
 &\qquad \qquad \qquad   \langle \W\rangle = e^{-\hat \G_1}  e^{-S_{_{\rm M2}}} \big[1 + \O({\T_2}^{-1})\big] \  , 
 \qquad \ \  \ \ \ \hat \G_1  =- \ha \sum_{n=-\infty}^\infty \log  X_n \ , \la{99}\\
&  X_n  =\frac{\big[{\rm det}(-\nabla^2+\frac{1}{4}R^{(2)} +{1\ov 4} (m_n +1)^2)\big]^{2}
\,
\big[{\rm det}(-\nabla^2+\frac{1}{4}R^{(2)}+{1\ov 4} (m_n-1)^2)\big]^{{2}}
\,
}{
\big[{\rm det}\big(-\nabla^2+1 + m_n  +\k^2 n^2  \big) \big]^{1\ov 2}  \,
\big[{\rm det}\big(-\nabla^2+1 -  m_n  +\k^2 n^2  \big) \big]^{1\ov 2} \, {\big[\rm det}\big(-\nabla^2+ \k^2 n^2 \big)\big]^3
      }, \no\\  
      \ \ \ 
&
 \ \ \ \ \ \qquad \ \  \qquad \qquad  \ \  m_n \equiv  \sqrt{1 + 4 \k^2n^2} \ . \la{31} 
\end{align}
The $n=0$  part here is the same as the  AdS$_3 \times S^3 \times T^4$  string contribution  ($\n=3$)  in 
\rf{3}. 

As a result, using  the $\zeta$-function regularization as in  \rf{12} we  conclude  
that the  log divergent part  of $\hat \G_1$  again   vanishes as in  \rf{225} so that $\G_1$  is finite.

Assuming that $\k$   is large   enough\foot{This
condition is  somewhat  analogous   to $k > 2$   in  the AdS$_4$ case in \rf{011}.}
  ($\k^2 > {3\ov4}$ so that $m_n >2$)
 which, according to \rf{29} 
 is the case if   $v_4$ is small  or $Q_5$ is large  which  corresponds
  to  the perturbative type IIA string theory regime, we  will 
 find that 
 after some non-trivial  cancellations  
 (that are due to underlying extended  (4,4) 2d supersymmetry of the spectrum in \rf{31}, see sections 3  and 4)\foot{
 As discussed in Appendix \ref{apd} an 
   extended 2d supersymmetry  is also responsible 
 for the  simplicity of the result \rf{011}  in the AdS$_4$ case.} 
 $\hat \G_1$  in \rf{99}  is  given  simply  by (cf. \rf{11},\rf{12}) 
\be \la{32}
\hat \G_1 
=  - \ha { \log} \, (2\pi ) + \sum _{n=1}^\infty{\log}\, (\k^2n^2)= -\log \frac {\k}{{\sqrt{2\pi}}}\ . \ee
The first  ($n=0$) term   is the  same as the finite part of the  string theory  ($\n=3$) contribution in \rf{5}. 

For smaller values of $\k^2= { v_4^{-2} u^{-1}} Q_5  \leq { 3\ov 4}$ 
(corresponding to  non-perturbative  string theory or M-theory regime) 
 $\hat \G_1$  in \rf{32} 
 gets   modified  according  to \rf{3.66},\rf{3.65}  as discussed in section 4    below. 

Thus for $\k > \sqrt {3\ov4}$     the  1-loop  M2  brane partition function in the  \adsts\ case    is given by 
\begin{equation}
  \langle \W \rangle =e^{-S_{_{\rm M2}}   -  \hat \Gamma_1 + ...}  
  =\frac {\k}{\sqrt{2\pi}} \,  e^{-  S_{_{\rm M2}} }\Big[ 1 + \O(\T_2^{-1}) \Big]
  ={ \sqrt Q_5 \ov \sqrt {2\pi u}\, v_4}       \,          e^{{ v_4\ov \sqrt{\hv u} } \sqrt{{Q_2 Q_5}}   }\Big[ 1 + \O(\T_2^{-1}) \Big]
     \  ,\label{36}
\end{equation}
where $ \O(\T_2^{-1})$ stands for possible 2-loop  M2 brane contribution.

    As in the AdS$_4 \times S^7/\mathbb{Z}_k$  case, we  find again that the  M2 brane 
generalization ``regularises  away''  the 
 1-loop  divergence of the GS string \rf{4}   replacing it with the expected \ci{Giombi:2020mhz}  finite
  1-loop prefactor proportional to  $\sqrt{ \T\ov 2\pi}$   (cf.  \rf{44},\rf{6}).
Compared to the result in the  \adsss\ case \rf{11}  where $  \frac{1}{2\sin\frac{2\pi}{k}} = {k\ov 4 \pi} [1 + \OO({1\ov k^2}) ]$
 in \rf{36}  here we do not get any  subleading  $1/ \k$  corrections.

  The effective string tension and  the   ``near-horizon''  value $\g$ of the string coupling (dilaton) 
  in  the corresponding  AdS$_3 \times S^3 \times T^4$  type IIA   background  which is the
   limit of   the  D2-D4  solution  are given by 
   (here $Q_5=Q_4$,  see \rf{204} and Appendices \ref{apb} and  \ref{apc})
  \begin{align}
{\rm IIA}: \ \ \  & \T=  2 \pi v_4 \T_2   
  = 
  {1\ov 2\pi} \sqrt{ v_4 \ov  v_1 v_2 v_3  u }  \sqrt{Q_2 Q_5}  =   { \gs \ov 2\pi  \sqrt{ \vv \uu  }}  \sqrt{Q_2 Q_5}   \ ,  \la{155} \\
  & \qquad   S=S_{_{\rm M2}} = -2\pi \T = -{ \gs \ov  \sqrt{ \vv \uu  }}  \sqrt{Q_2 Q_5}   \ ,  \la{1555} \\
 &  \vv = \vv_1 \vv_2 \vv_3 \  , \qquad  \uu = \ls^{-1} \lpl u \ ,  \qquad \vv_i =  \ls^{-1} \lpl  v_i \ ,   \qquad  \ \ \ 
 \ls^{-1} \lpl = \gs^{\ftr} \ ,   \qquad \ \ \  i=1,2,3 \ , \la{2555}
  \\
  &  \qquad 
  \g^2=  
    \gs^2    \sqrt{ \uu\ov \vv } \sqrt{  {Q_2\ov  Q_5}}   \ , \qquad  \qquad  
    { \sqrt \T \ov  \g}  = { \k \ov \sqrt {2 \pi}  } = {1 \ov \sqrt{ 2 \pi  \gs  \uu}}\sqrt{ Q_5} \ .
     \la{55} \end{align}
Here  $\T= {L^2_s\ov 2\pi \ls^2}$     with $L_s^2 = \sqrt{\C_2\C_5}$   being the  \AdS$_3$ radius  in 
the type IIA metric in \rf{C.5}, $\g = e^{\phi}$   with 
 $\gs$  being  its  ``asymptotic''  value  and 
 we  have redefined the moduli  parameters so that  the corresponding  $T^3 \times S^1$  radii
 in the 10d metric 
are  ``measured'' in units of $\ls$. 
Then   \rf{36} takes the form (cf. \rf{204}) 
\begin{equation}
  \langle \W \rangle ={ \sqrt \T \ov  \g}  \,   e^{2\pi \T } \Big[ 1 + \O(\T_2^{-1}) \Big] 
  \  ,\label{66}
\end{equation}
where the one-loop contribution $\kappa/\sqrt{2\pi}$ in \rf{32} is written in terms of the string parameters $\T$   and $\g$  as in   \rf{55}.
 Compared to the  AdS$_4 \times S^7/\mathbb{Z}_k$  case \rf{33},\rf{4444}, here,  surprisingly, 
 we do not get any  leading  large tension   corrections  $ (\g^2 \T^{-1})^p$  at  higher genus  orders.\foot{
 It is interesting to note that  a  somewhat similar   result  was   found in \ci{Gautason:2025bft}.
 This work   studied $1\ov 2$-BPS Wilson loop  for a   class  of $\N\geq 2$   3d  Chern-Simons-matter theories
  dual to AdS$_4 \times  {\rm SE}^7/\mathbb Z_k$ backgrounds.  
   In  certain cases  of ${\rm SE}^7/\mathbb Z_k$    spaces   it was   observed   that 
   1-loop AdS$_2 \times S^1$   M2-brane partition function  is   given just   by    the leading string theory  1-loop  contribution.
  }
 
 To summarize, the  M2  result  (\ref{66}) thus  reproduces  the type II string theory 
  one-loop expression  (\ref{6}) (with $\rm p=1$).
  The   latter had $1/\g$  factor included  to indicate the disk topology 
   and the $\sqrt{\T}$   factor  originated \cite{Giombi:2020mhz} from the replacement of the divergent 
    contribution in \rf{4}   by \rf{44} (which should be due to the contribution 
    of the GS  string path integral measure). 
   The M2   brane  computation (that includes  not just $n=0$ string modes 
   but also also all   higher $n$ modes   in \rf{32}) 
    automatically reproduces  this finite  ${ \sqrt \T \ov  \g} $ factor. 
 This is  the same what  happened  in the  $\ads_4 \times S^7/\mathbb Z_k$  case 
 (cf. \rf{011},\rf{11}):  in the large $k$  limit \rf{11}   gives 
 ${k\ov 4\pi}$  which  reproduces the prefactor in \rf{6} (with $\rm p=4$).

One may  speculate  that  further $\T_2^{-1}$ contributions  in \rf{36} may also be  absent
  so  that  the result for the $\ha$-BPS  WL 
in the AdS$_3 \times S^3 \times T^4$  string  case may be 1-loop  exact (like the central charge). 

 Expressed in terms of the parameters of the    \adst\   solution in type IIB string theory 
 which is  the near-horizon limit of the  D1-D5  background   T-dual  (along the $w$ circle in the 10d  limit of \rf{25})  
 to  the D2-D4  type IIA solution,  
  we get  instead   of \rf{55}
 
  \begin{align}
{\rm IIB}: \ \ \  \ \ \  & \T'  
  = {\gs'   \ov 2 \pi \sqrt{ \hat \vv' } } \sqrt{Q_2 Q_5}    \ ,   \ \qquad \ \ 
    \hat \vv'=  \vv\,  \uu^{-1} \ , \ \ \ \  \gs'  = \uu^{-1} \gs \ ,  \la{136}
  \\
  &  \g'=  
   { \gs'   \ov   \sqrt{\hat  \vv'}}  \sqrt{  {Q_2\ov  Q_5}}  
      \ , \qquad \qquad  
    { \sqrt{ \T'} \ov  \g'}  ={1\ov \sqrt {2 \pi}}   \sqrt[4]{{Q_5\ov Q_2}}\sqrt{Q_5}  \ , 
     \la{505}         \end{align}
 where  $Q_2=Q_1$ (see Appendix \ref{apc})  and  primes are  used to   indicate the  type IIB  parameters. 
 Under  the T-duality  $ \uu'=\uu^{-1}$  and $\gs'^{-2} \uu'=\gs^{-2} \uu $,   where 
 $\gs'$ is the type IIB  asymptotic   string coupling.
  We  also defined  $\hat \vv'=\vv \uu'$ as  the 
    dimensionless volume of $T^3\times S^1$.


 One  may  also   consider   an analog of  the  AdS$_2$  string minimal surface  in  \adst\   background   supported by a  ``mixed'' 
   combination of  RR and NSNS  fluxes. We  will discuss  the   corresponding 11d  background and find the AdS$_2 \times S^1$ 
    M2-brane  solution that generalizes the string  theory one in  \ci{Pajer:2021bfr}   in Appendix \ref{ape}. 
    The    1-loop correction  turns out  to have 
     the simple form \rf{32} with a  rescaled  parameter $\kappa$. 
     Similar result   is found  also  in the case of $\ads_3 \times S^3\times S^3 \times S^1 \times S^1$ background (see Appendix F). 
   
   There are several      generalizations of the present work. 
   One may    compute  the 2-loop    string  and also   2-loop      M2  brane corrections  to   the WL expectation value 
  following  the approach of  \ci{Beccaria:2025ahf,Beccaria:2026ffm}. 
 It would be   interesting also   use the semiclassical M2 brane  approach   to find  ``non-planar'' corrections  to 
anomalous dimensions of non-BPS operators 
 as  was done in   ABJM  context in  \ci{Giombi:2024itd,Beccaria:2025npl}.


\section{M2-M5 solution  and its \adsts\  limit}

Our   starting  point will be  
 the  $1\ov 4$-supersymmetric 11d supergravity  solution  representing 
  intersecting M2 and M5 branes 
   \cite{Tseytlin:1996bh} 
\begin{align}
ds_{11}^{2}=&H_{2}^{-\ftt}H_{5}^{-\ftr}\big(dx^2_{0}+dx^2_{1}\big)+H_{2}^{\ftr}H_{5}^{\ftt}\big(dr^2+r^2 ds_{S^{3}}^{2}\big)
+H_{2}^{\ftr}H_{5}^{-\ftr}d\y_p^2+H_{2}^{-\ftt}H_{5}^{\ftt}d\w^2, \label{2.1}\\
F_{4}=&dC_{3}=\big(i\, dH_{2}^{-1}\wedge dx_{0}\wedge dx_{1} -*_{4}dH_{5}\big) \wedge d\w, \label{2.2}\\
   H_{2}=&1+\frac{\C_{2}}{r^{2}},\qquad H_{5}=1+\frac{\C_{5}}{r^{2}}, \qquad\qquad   \C_{2}={\lpl^{2}\hv^{-1}\, Q_{2}}\ ,\qquad \qquad \C_{5}={\lpl^{2} u^{-1}\,  Q_{5}}
  \ . \label{2.4}
\end{align}
We consider  the Euclidean signature of the metric (see Appendix \ref{apb})   and assume  that the four 
``non-common''  
 $\y_p$  ($ p=1,2,3,4 $) directions  of M5 and   $\w$ direction  of M2 are  compactified on $T^4$ and $S^1$ respectively,
  i.e. 
\begin{equation}
 \y_p= \lpl y_p\ ,   \ \qquad   \ \  \w=\lpl w \ ,  \qquad 
  y_p \equiv y_p+2\pi v_p\ ,\qquad  w\equiv  w+2\pi u\  ,\qquad \hv\equiv v_1v_2 v_3  v_4 \ . \label{2.3}
\end{equation}
Here $\lpl$ is the 11d scale so that $y_p, w, v_p, u$ and $v$    are  dimensionless. The charge quantization condition 
 implies that $Q_2$ and $Q_5$ are integers.

In the  ``near-horizon'' limit, i.e.  dropping 1 from $H_2$  and $H_5$ in \rf{2.4},   we get the \adsts\ solution\foot{We shall use 
the notation $\vol(M)$  and $\vol_{M}$  respectively  for the volume and the volume form of space $M$.} 
\begin{align}
&ds^{2}_{11} =L^{2}\big( ds_{\rm AdS_3}^{2}+ds_{S^{3}}^{2}\big)  
+R^{2}dy^2_{p}+ R^{2}_w dw^2,\qquad\qquad  F_{4}={2L^{2} R_w}(i\text{vol}_{\AdS_3}+\text{vol}_{S^{3}})\wedge dw,\label{2.5}\\
& L^{2} =\big({\Q_2 \Q_5^2}\big)^{\frac{1}{3}} =   \lpl ^2\Big(\frac {Q_2 Q_5^2}{u ^2\hv }\Big)^{\ftr}, \ \  \qquad 
R^{2} = \lpl^2(\Q_2\Q_5^{-1})^{\ftr} = \lpl^2  \Big(\frac{Q_2 u}{Q_5 \hv }\Big)^{\ftr}  \ , \ \  \qquad 
R^2_w= \lpl^6 R^{-4} \ . 
\label{2.7} 
\end{align}
The  (Euclidean) AdS$_3$   metric    originates from \rf{2.1}   in  Poincare coordinates 
and can  then be transformed into global coordinates.  We  may rewrite \rf{2.5} as  ($i=1,2,3$) 
\begin{align}
&\la{277}  
ds^{2}=L^2\big(ds_{\rm AdS_3}^{2}+ds_{S^{3}}^{2}+dy'^2_{i}+dw'^2 + \k^{-2}d\x^2\big), \qquad \ \ \  
F_4=2L^3(i\text{vol}_{\AdS_3}+\text{vol}_{S^3})\wedge dw'\ ,   \\
&y'_i = L^{-1} R\,   y_i \ , \qquad w'= L^{-1} R_w\,   w \ ,  \qquad  \x= v_4^{-1} y_4 \equiv \x + 2\pi \ , \qquad  
\k^{-1}\equiv  L^{-1} R\,   v_4 \ ,
\label{3.5}\\
&ds_{\AdS_3}^{2}=\frac{(1+\frac{1}{4}z^2)^2}{(1-\frac{1}{4}z^2)^2}ds^2_{\AdS_2}  +  \frac{dz^2}{(1-\frac{1}{4}z^2)^2} \ , 
\qquad\qquad \ \  ds^2_{S^3}=\frac{dq^a dq^a}{(1+\frac{1}{4}q^a q^a)^2}\label{3.7} \ , \qquad \ \  a=1,2,3 \ , 
\end{align}
where AdS$_3$ and $S^3$  in \rf{3.7} have unit radii. 

The M2-M5  solution is related  by dimensional  reduction
 (along a $T^4$ direction  $y_4= v_4\, \x$) to  10d type IIA    D2-D4 solution.\foot{Reduction  along the $w$  direction leads to the  NS1-NS5 solution.} 
 It is  
   in turn T-dual (along $w$)   to the type IIB  D1-D5   solution    having  the  same numbers
    of respective branes or   integer charges 
      and  related values of moduli parameters   (see Appendix \ref{apc}).

\section{AdS$_2 \times S^1$  M2  brane        solution} 

Our aim  will be  to compute the  1-loop M2 brane   partition function 
expanded near the    AdS$_2\times S^1$ minimal  surface. 
We shall assume that $S^1$  corresponds to  $y_4=v_4\, \x$ direction  in \rf{2.3}. If $y_4$ 
  is identified with  the  11-th direction, 
 then the  ``double dimensional''   reduction    will    lead to  type IIA string  wrapped on AdS$_2$ 
    in   AdS$_3 \times S^3 \times T^3 \times S^1$  which is the near-horizon limit of the D2-D4  solution. 

 We shall  
consider 
 the  M2  brane  solution for which  two  of the 
   world volume   coordinates  $\s^1,\s^2$  are identified  with the AdS$_2$   coordinates
    inside  AdS$_3$ in \rf{3.7} 
   (with \AdS$_2$ located at $z=0$) 
and   the third one $\s^3\equiv \s^3 + 2\pi$    with the  angle $\vp$ in \rf{277}.  
One can check  that   such  M2 brane   configuration solves  the equations  of motion and 
is a  $\ha$-BPS one  (see Appendix \ref{apa}). 

The  value of the classical  M2 brane   action in \rf{A.1} is then   
\be \la{3.3} 
S_{_{\rm M2}}= T_2 \,  L^3 \,   ( 2\pi \k^{-1}) \,  \vol(\AdS_2) = 
 \T_2  \, ({ 2\pi v_4})\,  ( - 2\pi )  = -   { v_4\ov \sqrt{\hv u} } \sqrt{{Q_2 Q_5}} \ , 
\ee
  where  $\T_2 = L^2 R\, T_2 $ is the effective  dimensionless tension as  in (\ref{28}).
  Note that the potential  $C_3$   corresponding  to $F_4$ in \rf{2.5}   does not contribute to the 
  classical  value of the action.

  We shall consider fluctuations near this \AdS$_2\times S^1$   solution  fixing the static gauge as in, e.g., \ci{Giombi:2023vzu}. Representing the induced  3d  world-volume metric as  
  \begin{align}
    h_{mn}d\s^m d\s^n=\hat  h_{\hat m\hat n} d\s^{\hat m}d\s^{\hat n }+ h_{33} d\s^3d\s^3,\qquad  h_{33}= \k^{-2} \ , \qquad 
    m,n=1,2,3,\qquad \hat m,\hat n =1,2,\label{3.11}
    \end{align}
    where $\hat  h_{\hat m\hat n}$ is the unit-radius  \AdS$_2$ metric and $\s^3=\vp$ is $2\pi$ periodic,   we get from \rf{277},\rf{3.7} 
    for the ``volume''  part (cf. \rf{A.1}) of the  bosonic quadratic fluctuation   Lagrangian\foot{We  scale out  the overall factor of $\T_2$ in the quadratic action.}
    \begin{align}
    \mathcal{L}_{\rm V}^{(2)}=&\frac{1}{2}   \big( \partial^{{m}}z\partial_{{m}}z+  2 z^2  + 
   \partial^{{m}}q_{a}\partial_{{m}}q_{a}+\partial^{{m}} y'_i\partial_{{m}}y'_i   +  \partial^{{m}} w'\partial_{{m}}w'\big)
 \label{3.10}\ . 
\end{align}
Here  
$z$ and $q_a$ are fluctuations of  the coordinates in \rf{3.7}.  We thus  get one massive  and 3+3+1 massless  3d fields propagating on \AdS$_2\times S^1$. 

The contribution of the WZ term  to  the fluctuation part of the  M2 action 
 \rf{A11}  is found by choosing  the potential $C_3$  corresponding to $F_4$ in \rf{277} as 
$C_3=2L^3(i\text{vol}_{\AdS_3}+\text{vol}_{S^3})\, w'$  so that   
$ \mathcal{L}_{\rm WZ}^{(2)}= - 2\k\,  w'  \partial_{3} z+...$. Then  the total  quadratic 
bosonic  fluctuation  action  may be written as 
$\T_2 v_4 \int d^3 \s \, \sqrt{ \hat h} \, \big( \mathcal{L}_{\rm V}^{(2)} + \mathcal{L}_{\rm WZ}^{(2)}\big)$  where 
\begin{align}
   \mathcal{L}_{\rm B}=  \mathcal{L}_{\rm V}^{(2)} + \mathcal{L}_{\rm WZ}^{(2)}=
   \tfrac{1}{2} \Big[ & \partial^{{\hat m}}z\partial_{{\hat m}}z+ \k^{2}  (\del_3 z)^2 +  2 z^2  + 
   \partial^{{\hat m}} w'\partial_{{\hat m }}w'    + \k^{2}  (\del_3 w')^2 - 4\k\,  w'\, \partial_{3}z  \no \\  & +
   \partial^{{\hat m}}q_{a}\partial_{{\hat m }}q_{a}+   \k^{2}  (\del_3 q_a)^2  
    +     \partial^{{\hat m}} y'_i\partial_{{\hat m}}y'_i   +   \k^{2}  (\del_3 y'_i)^2  \Big]\ .   \la{3.16}
\end{align}
To diagonalise the $(z,w')$  part of the action 
we may expand  the 3d fluctuation  fields in  
Fourier modes   in  $\sigma^3$, thus replacing $\partial_3\rightarrow i n$,  $n=0, \pm 1, ...$. 
Then the $(z_n,w'_n)$ kinetic operator becomes 
\begin{align}
&  \bigg(\begin{array}{cc}
-\nabla^2+2+\k^2{n}^{2} & i\k\, n\\
-i\k\, n &-\nabla^2+ \k^2{n}^{2}
\end{array}\bigg)\label{3.17}\ \ \to \ \ \   \bigg(\begin{array}{cc}
-\nabla^2+ M^2_{n,+}  & 0  \\
0 &-\nabla^2+ M^2_{n,-}
\end{array}\bigg)\ , \\
& \qquad \ \ \ M^2_{n,\pm}
    =1 + \k^2 n^2 \pm  \sqrt{1+4 \k^2 n^2}\ , \qquad \qquad  
    n\in \mathbb{Z} \ . 
     \la{318}
\end{align}
where $\nabla^2$  is the AdS$_2$ Laplacian.  
 As a result, 
 the   bosonic part of the 1-loop M2  brane   partition function  is represented by the 
 contributions   of 8 towers of  bosonic modes in \AdS$_2$  
\be 
\hat Z_{1\rm B}= \prod_{n=0, \pm 1, ...}      \Big[\det(-\nabla^2+M^2_{n,+}  )\Big]^{1\ov 2}
    \Big[\det(-\nabla^2+ M^2_{n,-} )\Big]^{-{1\ov 2}} \Big[\det(-\nabla^2+ \k^2 n^2)\Big]^{-3}
 \ . \label{3.20}
\ee
The $n=0$ contribution here is the same as  the  10d string contribution  in \rf{3}  with  $\n=3$.


The  quadratic    fermionic  part of the M2 brane   action  in \rf{A.5}  may be written as\foot{Here we  formally  continue to the Euclidean signature 
so that  the resulting  fermionic  differential operator  can be defined on  Euclidean \AdS$_2$.}
 \begin{align}
  & S^{(2)}_{\rm F}= iT_2 \int d^3 \s \;\sqrt{ h} \, \bar \theta\,  \Gamma^m (1-\Gamma_\ka)\hat D_m \theta,\label{3.21}\\
  &\Gamma_\ka =\frac 1 {3!\sqrt h}\epsilon^{mnp}\Gamma_{mnp},\qquad \Gamma^m=h^{mn}\partial_n X^N\Gamma_N,\qquad \hat D_m=\partial_m X^M \hat D_M.\label{3.22}
\end{align}
The explicit form of the  spinor  derivatives is given  in Appendix \ref{apa}.
 Using underlined indices  for tangent  space components of Dirac matrices  $\Gamma^A$
 and  labelling $\vp$ in \rf{277}  as the 9-th direction we  observe that 
$\Gamma_\k=\Gamma^{\underline{019}},\;[\Gamma_\ka,\hat D_m]=0$, so that 
the Lagrangian  in \rf{3.21} 
can be written as $ \bar \theta \Gamma^m (1-\Gamma_\ka)\hat D_m \theta= \bar \theta \Gamma^m \hat D_m [(1-\Gamma_\ka)\theta]$. We fix  $\ka$-symmetry gauge  by the condition 
 $\tfrac 12 (1-\Gamma_\ka)\theta=\theta$. Then   
\begin{align}
    S_{\rm F}^{(2)}= 2iT_2 \int d^3 \s \, \sqrt{ h}\,  \bar \theta\,  \Gamma^m \hat D_m \theta\label{3.27}
\sim  \int d^3 \sigma \,\sqrt{\hat  h }\, \Big[\bar \theta  \Gamma^m \nabla_m \theta 
-\tfrac i{2} \bar \theta\Gamma^{\underline {012\,10}} (1-i\Gamma^{\underline{012345}})\theta\Big]\ , 
\end{align}
where $ \Gamma^m \nabla_m  = \Gamma^{\hat m} \nabla_{\hat m} + \Gamma^{3} \del_3$  corresponds to the Dirac operator for the AdS$_2\times S^1$ metric in \rf{3.11}.
As in the \adsss\  case \cite{Sakaguchi:2010dg}, expanding in Fourier modes  in $\s^3$  
 the resulting  fermionic Lagrangian  may be  written in terms of 
    towers  of complex 2d fermions $\psi^r_n,\,\chi^r_n$ ($r=1,2$)    propagating on  AdS$_2$
\begin{align}
      \mathcal{L}^{(2)}_{\rm F}=\sum_{n=0, \pm 1, ...} \Big[ \bar{\psi}^{r}_{-n}&
      \rho^{\hat{m}}\nabla_{\hat{m}}\psi^{r}_n  +\bar{\chi}^{r}_{-n} \rho^{\hat{m}}\nabla_{\hat{m}}\chi^{r}_n
      +\tfrac{1}{2}\mu_{-}\bar{\psi}^{r}_{-n} \rho_{{2}}\psi^{r}_n 
      +\tfrac{1}{2}\mu_{+}\bar{\chi}^{r}_{-n} \rho_{{2}}\chi^{r}_n\no \\ 
      & -\tfrac{1}{2} \bar{\psi}^{r}_{-n}\chi^{r}_n -  \tfrac{1}{2} \bar{\chi}^{r}_{-n}\psi^{r}_n \Big]\ , 
      \qquad 
      \qquad   \mu_{\pm }=\pm 1-2\k\,n\  .\label{3.31}
\end{align}
Here $\rho^{0},\rho^{1}=(\hat \sigma_2, \hat \sigma_1)$  and $ \rho^{{2}}=\hat \sigma_3
$ are 3d  Dirac  matrices ($\hat \s_i$ are Pauli matrices).

Computing the  fermionic determinant 
 corresponding to \rf{3.31} we get  the following  contribution to the partition function (cf. \rf{318},\rf{3.20}) 
\begin{align}
   \hat Z_{1\rm F} = &\prod_{n=0, \pm 1, ...}   \big[\text{det}\big(-\nabla^2+\tfrac 14 \RR+ \M^2_{n,+} \big)\big]^2
\big[\text{det}\big(-\nabla^2+\tfrac 14 \RR+  \M^2_{n,-}  \big)\big]^2\label{3.32} \ , \\
& \M_{n,+}
    =  \tfrac 12 \big(\sqrt{1 +   4 \k^2 n^2} +1  \big)   \ , \qquad   \qquad \ \ \ \M_{n,-}
    =  \tfrac 12 \big(  \sqrt{1 +   4 \k^2 n^2} -1  \big)  \ , \la{3.33} 
\end{align}
where $-\nabla^2 +  \tfrac 14 \RR $   is  the square of the Dirac operator on AdS$_2$   ($\RR=-2$). 
The  $n=0$  contribution  (with $ \M^2_{0,\pm}=1$)  in \rf{3.32} 
 is the same as in the $\n=3$  case of the string partition function in \rf{3}. 

Combining \rf{3.20}  and \rf{3.32}  we  get  for the 1-loop M2 brane partition function 
\be 
\hat Z_1 = \prod_{n=0, \pm 1, ...} {   \big[\text{det}\big(-\nabla^2+\tfrac 14 \RR+ \M^2_{n,+} \big)\big]^2
\big[\text{det}\big(-\nabla^2+\tfrac 14 \RR+  \M^2_{n,-}  \big)\big]^2  \ov 
 \big[\det(-\nabla^2+M^2_{n,+}  )\big]^{1\ov 2}
    \big[\det(-\nabla^2+ M^2_{n,-} )\big]^{1\ov 2} \big[\det(-\nabla^2+ \k^2 n^2)\big]^{3}}
\label{4.42} \ . \ee
Note  due to  the residual AdS$_2$   supersymmetry that should be present  
 in the quadratic fluctuation spectrum after fixing the static and $\ka$-symmetry gauges (cf. \ci{Sakai:1984vm,Drukker:2000ep,Sakaguchi:2010dg,Giombi:2023vzu}) 
  the  bosonic and fermionic modes  for each  level $n\not=0$  should form   long massive 
   $\N=$(4,4)     supermultiplets  of $\mathfrak{psu}(1,1|2)$  with the  bosonic and fermionic 
   masses  of the 8+8  modes  related by  (see Appendix \ref{apd})\foot{\la{ff1}Note 
   that  for  bosons and fermions     belonging to  $\N=(1,1)$  AdS$_2$
    supermultiplets   one has  $m_{\rm B}^2 = \Delta(\Delta-1), \  m_{\rm F} =\Delta $ so that 
    $m_{\rm B}^2= m_{\rm F}^2 - m_{\rm F}$.}
 \begin{align}
&\;m_{\rm B}^2:\,\qquad \Delta(\Delta-1)\; \ \ \text{(1 mode)};\qquad \;(\Delta+1)\Delta\;\ \  \text{(6 modes)}; \qquad \;(\Delta+2)(\Delta+1)\;\ \  \text{(1 mode)}\ , \la{315}\\
&\;|m_{\rm F}|:\; \qquad \Delta\;\ \ \ \ \ \ \text{(4 modes)}; \qquad \;\qquad\;\;\;\qquad\Delta+1\;\ \ \text{(4 modes).}\la{316}
\end{align}
These  relations are indeed satisfied for the spectrum in \rf{4.42} 
  for  $\Delta$ equal to  
    \al{ 
&  \Delta_n =    \tfrac 12(m_n-1)\ , 
 \qquad \ \ \ \ \  m_n  \equiv \sqrt{1+4\k^2 n^2}   \ .  \la{317} 
 }
The bosonic modes  have 1d scaling   dimensions   $1\times \Delta_n$, $6\times (\Delta_n+1)$, $1\times (\Delta_n+2)$.
The  condition  $\Delta_n \geq  {1\ov 2}$   
which  corresponds to the  choice of  $\Delta=\Delta_+ = \ha (1 + \sqrt{1 + 4 m_{\rm B}^2})$
  branch  above the BF threshold  (i.e.  the Dirichlet  boundary conditions as appropriate for the 2d fluctuation fields)
  is satisfied  provided $m_n \geq 2$.  This 
   applies  for all $n^2 > 0$ if $\k^2 \geq {3\ov 4}$.

A check of  consistency  of the  above spectrum with the AdS$_2$   supersymmetry  is the vanishing of the total  AdS$_2$ vacuum energy \ci{Drukker:2000ep,Giombi:2023vzu}.  Using that $E_{\rm B}=-\frac 14 (m_{\rm B}^2+\frac 16),\ \ 
 E_{\rm F}=\frac 14(m_{\rm F}^2-\frac 1 {12})$   
 we get  for the spectrum in \rf{4.42}   (for each value of $n=0, \pm  1, ...$) 
  \be\la{444}
   \te  E^{(n)} _{\text{tot}}=-\frac 14\Big[ M^2_{n,+} + M^2_{n,-} + 6 \k^2 n^2  - 4 \M^2_{n,+}  - 4  \M^2_{n,-}  + 8\times  ( {1\ov 6}  + {1\ov 12} ) \Big] =0  \ . \ee
  
 \def \rB {{\rm B}} \def \rF {{\rm F}}
  
\section{1-loop  correction to M2 brane partition function} 

Starting with $\hat Z_1$  in \rf{4.42}  we  find  (cf. \rf{4},\rf{225}) 
\begin{align}
&\hat \G_1 = - \te \ha \log \hat Z_1 = \hat \G_{1\infty}  + \bar \G_1 \ , \ \ \la{410}  \ \ \ \\
& \hat \G_{1\infty}  = -\sum_{n=-\infty}^\infty B^{(n)}_2 \log\Lambda,\qquad\qquad 
 B^{(n)}_2=\tfrac 1{4\pi}\int d^2\sigma \sqrt {\hat h}\;b^{(n)}_2 \ , \la{420}
\end{align}
where the Seeley   coefficient $b_2$   for a 2d  operator $-\nabla^2 +  X$ is given by $b_2 ={\tr} ( \frac{1}{6} \RR - X)$. 
Using that $\vol(\AdS_2)=-2\pi$   and $\RR=-2$   we then get   from  \rf{4.42}  (cf. \rf{444}) 
\be\la{345}
\te     b^{(n)}_{2,\text{tot}} =- \Big[   M^2_{n,+} +  M^2_{n,-} + 6 \k^2 n^2 
 -4 \M^2_{n,+}   - 4   \M^2_{n,-}  - 8 \times {1\ov 4}   \RR  \Big] = -2   \ , \qquad \qquad 
  B^{(n)}_{2,\text{tot}} =1 \ . 
   \ee
Thus  for each $n$  the contribution  of the bosonic and fermionic modes  to the logarithmic 
divergence  has   the same   universal value as  in the string ($n=0$) case \cite{Giombi:2020mhz}
and as in the M2 brane  in  \adsss\   case \ci{Giombi:2023vzu}.  Summing over $n$ 
we  then get as in \ci{Giombi:2023vzu} that $B_{2,\text{tot}} =\zeta_{\rm tot}(0)=0$, i.e.\foot{The use of Riemann zeta function regularization  is consistent with the expected  absence of 1-loop 
 logarithmic  divergences in a 3d theory and is part of the definition of the M2  brane path integral 
  (cf. standard quantum mechanics path integral).} 
\be 
\hat \G_{1\infty} = -\log \Lambda \sum _{n=-\infty}^\infty1=-\log\Lambda\, \big[1+2\zeta_R(0)\big]
=0\ .\label{3.47}
\end{equation}
Using as in \ci{Drukker:2000ep,Buchbinder:2014nia,Giombi:2020mhz} 
the AdS$_2$  spectral  zeta-function regularization  to compute 
finite parts of  log det's of differential operators in  \rf{4.42} we get 
\begin{align}
     &\hat\Gamma_{1}=-\tfrac 12 \sum _{n=-\infty}^\infty
     \zeta'_{\text{tot},\,n}(0)\ , \label{3.48}\\
     \zeta'_{\text{tot},\,n}  (0)=&\zeta'_\rB(0; M^2_{n,+})+\zeta'_\rB(0; M^2_{n,-})+  6\zeta'_\rB(0; \k^2 n^2 )
     +4\zeta'_\rF(0;  \M_{n,+}   )+4\zeta'_\rF(0; \M_{n,-} )\label{3.49},
\\
    &\zeta_{\rB}^{\prime} (0;m_{\rm B}^{2})=-\tfrac{1}{12}(1+\log2)+\log {A}-\int_{0}^{m_{\rm B}^{2}+\frac{1}{4}}dx\ \psi\big(\sqrt{x}+\tfrac{1}{2}\big),\label{3.50}\\
     &\zeta_{\rF}^{\prime} (0;m_{\rm F})=-\tfrac{1}{6}+2\log {A}+\big|m_{\rm F}\big|+\int_{0}^{m_{\rm F}^{2}}dx\ \psi(\sqrt{x}) \ .\label{3.51}
\end{align}
 where $ A$ is  Glaisher's constant. 
 The  integrals of $\psi(x)= {d\ov d x} \log \Gamma(x)$  can be written (after 
  integration by parts) in terms of  Barnes  function $\GG(x)$   (see, e.g., \cite{Olver2010NIST}):
  \begin{align}
    &\int_{0}^{m_{\rm B}^{2}+\frac{1}{4}}dx\ \psi\big(\sqrt{x}+\tfrac{1}{2}\big)=\te 2\sqrt{m_{\rm B}^{2}+\tfrac{1}{4}} \log \Gamma\big(\sqrt{m_{\rm B}^{2}+\tfrac{1}{4}} +\ha  \big)+\frac 5{12}\log 2 \no 
    \\  
    &\qquad \qquad \qquad \qquad \qquad \qquad +\tfrac 12\log \pi+3\log A-2\psi^{(-2)}\big(\sqrt{m_{\rm B}^{2}+\tfrac{1}{4}}+\tfrac 12\big)\ , \label{3.53}\\
    &\int_{0}^{m_{\rm F}}dx\ \psi(\sqrt{x})=2|m_{\rm F}|\log \Gamma(|m_{\rm F}|)-2\psi^{(-2)}(|m_{\rm F}|)\ ,\label{3.54}\\&  \psi^{(-2)}(z)=\int _0 ^z dx\, \log \Gamma(x)
    =\tfrac{1}{2} z(1-z)+ \tfrac{1}{2} z \log (2 \pi)+(z-1) \log \Gamma(z)-\log \GG(z)\ . \label{3.52}
\end{align}

Using  the   notation  $  m_n= \sqrt{1 +  4 \k^2 n^2}  $ 
and considering the case of $m_n > 2$ 
  which  applies   for all $n^2 >0$ 
   if 
$\k^2 > {3\ov 4}$ we may express  
all $\zeta'(0)$-values  in \rf{3.49}  in terms of  $m_n$  as\foot{In general, the $\zeta'(0)$-values   contain terms 
 like $\log \G( {1 + |m_n-2| \ov 2})$  that are sensitive  to  the sign of $m_n-2$.    The $n=0$ case   corresponding to $m_0=1$  in  \rf{317}  can be also  treated separately (cf. \rf{5}).} 
    \begin{align}
    \zeta'_\rB(0;\k^2n^2)=&\te -\frac 1{12}+\frac 12m_n{\log}\, (2\pi)-2\log A-\frac 14 (m^2_n-1)\te -\log\frac{m_n-1}{2}-3\log\Gamma(\frac{m_n-1}{2})-2\log \GG(\frac{m_n-1}{2})\no , \\
    \zeta'_\rB(0; M^2_{n,-})=&\te -\frac 1{12}+\frac 12(m_n-2){\log}\, (2\pi)-2\log A-\frac 14(m_n-3)(m_n-1) 
    -\log \Gamma(\frac{m_n-1}{2})-2\log \GG(\frac{m_n-1}{2}), \no \\ 
     \zeta'_\rB(0;M^2_{n,+})=&\te -\frac 1{12}+\frac 12(m_n+2){\log}\, (2\pi)-2\log A-\frac 14(m_n+3)(m_n+1)-\log\frac{m_n +1}{2} ,\nonumber\\
    &\te  -3\log\frac{m_n-1}{2} -5\log\Gamma(\frac{m_n-1}{2})-2\log \GG(\frac{m_n-1}{2}),  \label{3.57}\\
\zeta'_\rF(0;  \M_{n,-})=&\te -\frac 16-\frac{m_n-1}{2}{\log}\, (2\pi)+2\log A+\frac 14 (m_n-1)^2 +2\log\Gamma(\frac{m_n-1}{2})   +2\log \GG(\frac{m_n-1}{2}), \no \\ 
    \zeta'_\rF(0; \M_{n,+})=&\te -\frac 16 -\frac{m_n+1}{2}{\log}\, (2\pi)+2\log A+\frac 14 (m_n +1)^2 \te +2\log \frac{m_n-1}{2} +4\log\Gamma (\frac{m_n-1}{2})+2\log \GG(\frac{m_n-1}{2})  \no 
\end{align}
where we   used that  $ \Gamma(y+1)=y\Gamma(y),\; \GG(y+1)=\Gamma(y)\, \GG(y)$.

Substituting these expressions into \rf{3.49}  we observe 
 some remarkable cancellations  so that the final result for $\k^2>{3\ov 4} $   is just 
\begin{align}
    &  \zeta'_{\text{tot},\,n=0}(0)=\log\, (2\pi) \ , \qquad \qquad  
       \zeta'_{\text{tot},\,n\not=0}(0)=-\log \big[\tfrac{1}{4} (m ^2_n-1)\big]=- {\log}\, (n^2 \k^2)
      \ . \label{3.61}
\end{align}
Thus   \rf{3.48} takes a very simple  form (cf. \rf{12}) 
\begin{align}
    &\hat \Gamma_{1}= -\tfrac 12 {\log}\, (2\pi) + \sum _{n=1}^\infty{\log} (n^2\k^2)=-\log \frac {\k}{{\sqrt{2\pi}}}\ , \label{3.63}
\end{align}
where we used again the Riemann zeta-function regularization (cf. \rf{12}). 
 The 1-loop partition function is then  given  by the same  expression   as  in the string  theory  case  (cf. \rf{6},\rf{66}) 
\begin{equation}
    \hat Z_1=e^{-\hat \Gamma_1}=\frac {\k}{\sqrt{2\pi}}\ .\label{3.64}
\end{equation}
The reason   for the cancellations leading  to \rf{3.61}    can be traced back  to  the (4,4) supersymmetry 
of the   mass spectrum in \rf{4.42}--\rf{317}.
 Indeed, considering the contribution of 
a   single  (4,4)  supermultiplet   in \rf{315}--\rf{316}  we  find  for its   1-loop contribution  (cf. \rf{3.49})\foot{Here we use  again  \rf{3.50}--\rf{3.52}  and $\psi^{(-2)}(z+1)=\psi^{(-2)}(z)-z+\frac 12\log 2\pi+z\log z$ and $
\Gamma(z+1)=z\Gamma(z)$.}  
\al{
\zeta'_{\text{tot}, \Delta}  (0) =&\zeta'_\rB(0; \Delta (\Delta-1))+  6\zeta'_\rB(0; (\Delta+1)\Delta ) 
+\zeta'_\rB(0; (\Delta +2) (\Delta+1))
     +4\zeta'_\rF(0;  \Delta   )+4\zeta'_\rF(0; \Delta +1  ) \no \\ 
=&2+8\Delta -4\log 2\pi +(6\Delta +1)\log \Gamma(\Delta)+(2-4\Delta)\log \Gamma(\Delta+1)
-(2\Delta+3)\log \Gamma (\Delta+2)\no \\
&-6\psi^{(-2)}(\Delta)+4\psi^{(-2)}(\Delta+1)+2\psi^{(-2)}(\Delta+2)=  -\log(\Delta (\Delta+1))\ .
\la{3344}
}
where   we assumed that $\Delta > {1\ov 2}$.  
This result   is equivalent to \rf{3.61}  if we set $\Delta$ equal to $\Delta_n$   in \rf{317}.\foot{Note   that in the case of $\N=(1,1)$ multiplet of one 
boson  and one fermion in AdS$_2$ (cf. footnote \ref{ff1}) one finds

$\zeta'_\rB(0;\Delta(\Delta-1))=-\frac 1{12}-\frac 12 \log 2\pi-2\log A-(2\Delta-1)\log \Gamma(\Delta)+2\psi^{(-2)}(\Delta)-2\Delta-\log\Delta$, 

$\zeta'_\rF(0;\Delta)=-\frac 16 +2 \log A+\Delta +2\Delta \log \Gamma(\Delta)-2\psi^{(-2)}(\Delta)$  \ \ so 
that the analog of \rf{3344}  

$\zeta'_\rB(0; \Delta (\Delta-1))  + \zeta'_\rF(0;  \Delta   ) 
 =-\frac{1}{4}+\Delta+\frac{1}{2} \log 2 \pi- \log \Gamma(\Delta)-\log \Delta$\ \ 
 still contains $\log \Gamma(\Delta)$  term. 
}

Thus   the simplicity of the result in \rf{3.61},\rf{3.3}   is due to  extended $\mathfrak{psu}(1,1|2)$   AdS$_2$ supersymmetry  
of the world-volume  action. 
Similar   observation  applies (though in a more non-trivial way) to the case of  AdS$_2\times S^1$   M2   brane in 
 AdS$_4 \times S^7/\mathbb{Z}_k$  considered  in \ci{Giombi:2023vzu}. As we   discuss in Appendix \ref{apd}
 there  the  non-trivial   cancellations in the 1-loop effective action in \rf{9},\rf{10}  leading to \rf{011}   can be understood as a consequence  of underlying 
  $\mathfrak{su}(1,1|3)$  supersymmetry.

\begin{figure}[h]
    \centering
    \includegraphics[width=0.5\linewidth]{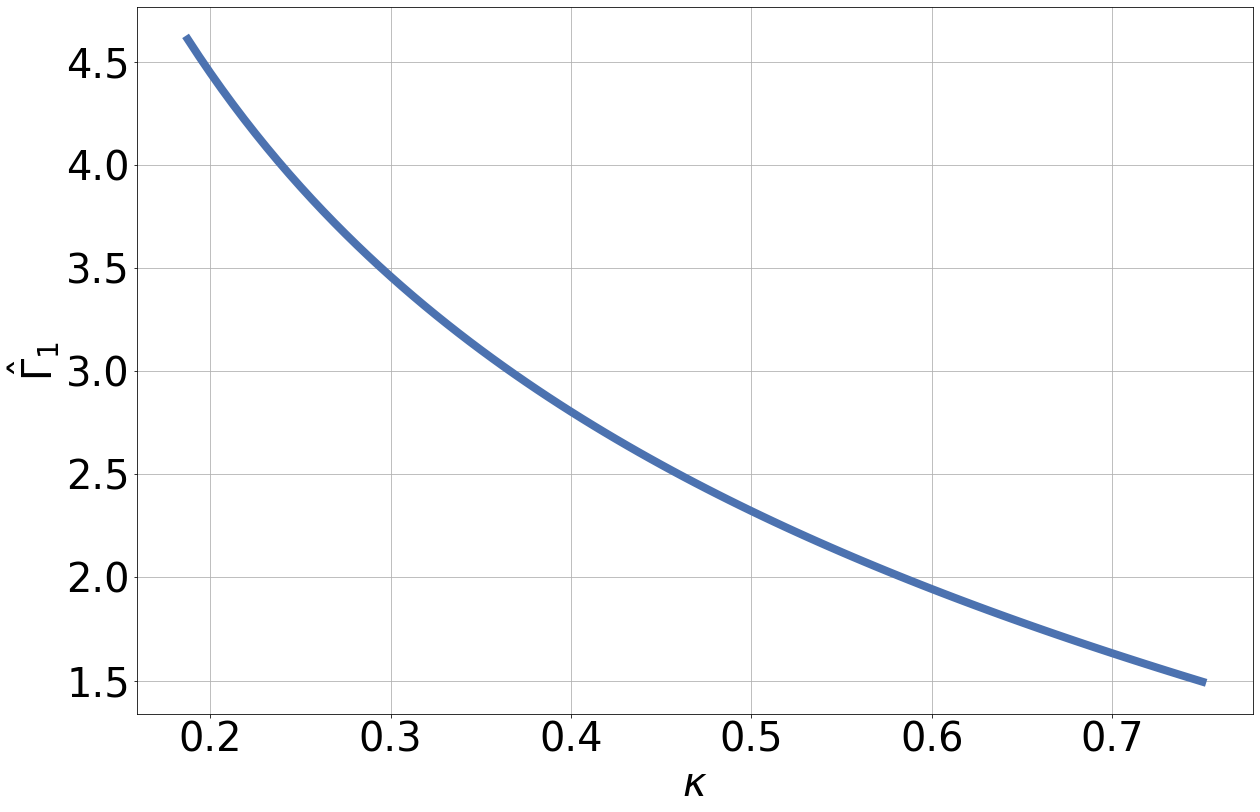}
    \caption{ $\hat \Gamma_{1} (\k)$  in the interval  ${\sqrt 3\ov 4}<\k< {\sqrt 3\ov 2} $.}
    \label{fig:1}
\end{figure}

When  
  $\k^2<{3\ov 4} $ 
   we find that  $\hat \Gamma_1$ 
    in \rf{3.48}
   contains  
   a   finite number  of terms   with  $|n|=1,..., n_*$ (where $n_*=n_*(\k)$)  such that $m_n <2$. 
  Then   instead of \rf{3.61} we  get\foot{This case is somewhat  analogous to the special cases of 
  $k=1,2$   values in the AdS$_4$ case  \ci{Giombi:2023vzu}.}   
 \begin{align}
    \zeta'_{\text{tot},\,n\leq n_*}(0)\equiv \z_n =(2-m_n){\log}\, (2\pi)-2\log\frac {\GG(\frac{1+ m_n}{2})} {\GG(\frac{3- m_n}{2})}-\log\Big[\Gamma(\tfrac{3+ m_n}{2})\Gamma(\tfrac{3- m_n}{2})\Big]\label{3.65}\ . 
\end{align}
Then    (cf.  \rf{3.63})
\begin{equation}
    \hat \Gamma_{1}=-\tfrac 12 {\log}\, (2\pi) -  \sum_{n=1}^{n_*} \Big[\tfrac{1}{2}  \z_n + {\log}\,(n^2 \k^2)\Big]
     +\sum_{n=1}^\infty{\log}\,(n^2 \k^2)
    =  - \log {\k\ov \sqrt {2 \pi}}   -  \sum_{n=1}^{n_*} \Big[\ha \z_n + {\log}\,(n^2 \k^2)\Big] \ .    \label{3.66}
\end{equation}
The  plot   of  $\hat \G_1$ in \rf{3.66}   as a function of $\k$ in the  special case of  $\k \in ({\sqrt 3\ov 4},{\sqrt 3\ov 2}) $   when $n_*=1$     is 
 in Fig. \ref{fig:1}.   

We  conclude   that while    for large enough  value of the parameter $\k$  in \rf{3.5} or \rf{29} 
(corresponding, e.g.,  to  a small  value of $v_4= \gs^{\ftt}$, i.e.  to  the  perturbative  type IIA  string regime) 
the  1-loop M2 brane partition function is  given by \rf{3.64}, i.e. it  does not  contain   subleading $\k^{-1}$ corrections, 
for smaller    values of $\k< {\sqrt{3}\ov 2}$ (corresponding to a  genuine M-theory regime) 
 it is   given by \rf{3.66}. Similar results  are found in the mixed flux case as discussed in Appendix E.

\section*{Acknowledgements}
AAT is grateful to M. Beccaria, S. Frolov,  B. Stefanski  and J. van Muiden   for  useful  discussions. 
ZW is grateful to K. Naderi and D. Zhong for discussions. 
We also thank J. van Muiden for valuable comments on the draft.
This work was supported by the STFC grant ST/T000791/1.

\appendix

\section{11d   supergravity  action  and  reduction  to 10d  
\label{apb} }

The  bosonic part of the action of 11d supergravity is given by 
\begin{align}
& S_{11} =\frac{1}{2\kappa_{11}^{2}}\Big[ \int d^{11}x\sqrt{-G}\big(R-\tfrac{1}{2}\big|F_{4}\big|^{2}\big)-\tfrac{1}{6}\int C_{3}\wedge F_{4}\wedge F_{4}\Big] ,\label{B.1}\\
& F_4= d C_3 \ , \qquad  \big|F_{4}\big|^{2}=\frac{1}{4!}F_{MNPQ}F^{MNPQ},\;\qquad \qquad  2\kappa_{11}^{2}=(2\pi)^{8}\lpl^{9} \ .\label{B.2}
\end{align}
The 11d metric  of  $M^{10} \times S^1$  may be written as 
\begin{equation}
ds_{11}^{2}=e^{-\frac{2}{3}\tphi}ds_{10}^{2}+e^{\frac{4}{3}\tphi}\big(dx^{11}+C_{\mu}dx^{\mu}\big)^{2},\qquad \qquad x^{11}\equiv  x^{11}+2\pi R_{11},\label{B.3}
\end{equation}
where $ds_{10}^{2}$  is type IIA string-frame metric   and 
 ${\tphi}=\phi-\phi_0$   where   $\phi_0=\log \gs$ is the asymptotic  (large-distance)  value of the  dilaton 
   so  that in the  flat space limit  both 
 $ds_{11}^{2}$  and  $ds_{10}^{2}$      have  the  standard  Minkowski form.
Then  the relations  between $R_{11}$ and  the asymptotic string coupling $\gs$ 
and also  the  string and membrane tensions  are\foot{Alternatively, one may
  replace  $\tphi$ in \rf{B.3} by the  full $\phi$ and then do constant  rescaling of   the 11d coordinates to put the  11d metric in  the canonical flat  space form. In 
  general, the  vacuum expectation value of the metric   may be chosen as $c_0 \eta_{MN}$
 where $c_0$ will be related to $\phi_0$ if the 10d metric has canonical form.
 In that case $R_{11}= \lpl =\ell_s$ (see  e.g.  Appendix A in \ci{Beccaria:2023hhi}).} 
\begin{align}
&R_{11}=\gs \ls= \gs^{\ftt} \lpl,\qquad \qquad\lpl =\gs ^{\ftr}\ls\label{B.6}\ ,\qquad \qquad 
\frac{2\pi R_{11}}{2\kappa_{11}^{2}}=\frac{1}{ 
  2\kappa^{2}_0 g_{s}^{2}}\ , \qquad \ \  2\kappa^{2}_0 \equiv  (2\pi)^7 \ls^8\ , 
 \\
&\qquad \qquad T =\frac{1}{2\pi\ls^{2}} \ , \qquad \qquad   T_2 = \frac{1}{(2\pi)^{2}\lpl^{3}}\ , \qquad\qquad
T= 2 \pi R_{11} T_2 \ . \label{B.5}
\end{align}
The resulting    10d   type IIA supergravity action 
  may be written as 
  \begin{align}
  S_{10A}=&\frac{1}{2\kappa^{2}_0  \gs^2 } \int d^{10}x\sqrt{-G}\Big[ e^{-2\tphi}\big(R+4(\partial\tphi)^{2}-\tfrac{1}{2}\big|\H_{3}\big|^{2}\big) -\tfrac{1}{2} \big(\big|{F}_{2}\big|+\big|F_{4}\big|^{2}\big)+...\Big] \no  \\
  =&\frac{1}{2\kappa^{2}_0   } \int d^{10}x\sqrt{-G}\Big[ e^{-2\phi}\big(R+4(\partial\tphi)^{2}-\tfrac{1}{2}\big|\H_{3}\big|^{2}\big)
   -\tfrac{1}{2} \big(\big|{\FF}_{2}\big|+\big|\FF_{4}\big|^{2}\big)+...\Big]  \label{B.7}\ , 
\end{align}
where  $ e^\phi=  \gs e^\tphi$   and $\FF_{p} = \gs^{-1}  F_{p}$.

In  section  2 we  consider  the  M2-M5  11d solution \rf{2.1}  with 5  isometric  toroidal directions.
We  use the following notation for  the coordinates of  $T^5$  with  flat  asymptotic metric\foot{For notational simplicity we will not distinguish between $\y_p$ and $\y^p$, etc.} 
\al{
&  \qquad \qquad ds^2_{T^5} =  d\y_p d\y_p  +  d\w^2  \ ,\la{a11}\\
&   \y_p= \lpl y_p\equiv \y_p + 2\pi \lpl v_p \ ,   \ \qquad 
  \ \  \w\equiv \w + 2 \pi \lpl u \ , \ \ \  \qquad p=(i,4), \ \  i=1,2,3 \ ,  \la{a88} \\
&
  \y_p= \lpl y_p \ , \ \ \  \w=\lpl w \ , \ \ \ 
    y_p \equiv y_p+2\pi v_p\ ,\qquad  w\equiv  w+2\pi u\ , \qquad 
    \ \  \hv\equiv v_1v_2 v_3  v_4  
     \ .  \label{a778}
}
In reducing to 10d we   shall  identify $\y_4$   direction with  $x^{11}$ in \rf{B.3}    so that  according to \rf{B.6} 
\be 
R_{11}   = \lpl v_4 \ ,  
\qquad  \qquad    \ \  v_4  = \gs^{\ftt} \ .\la{a77}  \ee
We shall also define the following dimensionless parameters 
\al{
&\la{a222}\vv_i \equiv  \ls^{-1} \lpl v_i = \gs^{\ftr} v_i \ , \ \ \ \ \ \ \ \ \ \  \vv_4 =  \ls^{-1} \lpl v_4=  \gs \ , \qquad \ \ \    \uu\equiv  \ls^{-1} \lpl  u  = \gs^{\ftr}  u\ ,  \\
&\hv= v_1v_2 v_3  v_4 = \gs^{-\ftr} \,\vv  \ , \qquad \ \ \  \ \   \vv\equiv \vv_1 \vv_2 \vv_3 \ . \la{a00}}

\section{D2-D4 and  D1-D5    solutions
\label{apc} }


The dimensional reduction  of  the  M2-M5 solution  \rf{2.1},\rf{2.2}  on
$\mathrm y_{4}$ direction gives the  D2-D4 solution of  IIA  supergravity  action \rf{B.7} that can be written as    \cite{Tseytlin:1996bh}
  ($i=1,2,3$; $\ls=\sqrt{\alpha'}$) 
\begin{align}
  &ds_{10A}^{2}=H_{2}^{-1/2}H_{4}^{-1/2}\big(dx_{0}^{2}+dx_{1}^{2}\big)+H_{2}^{1/2}H_{4}^{1/2}\big(dr^{2}+r^{2}ds^2_{S^3}\big)+H_{2}^{1/2}H_{4}^{-1/2}d\mathrm{y}_{i}^{2}+H_{2}^{-1/2}H_{4}^{1/2}d\mathrm w^{2},\label{C.1}\\
 &\ \FF_{4}= g_{s}^{-1}\big(i\, dH_{2}^{-1}\wedge dx_{0}\wedge dx_{1}-*_{4}dH_{4}\big)\wedge d\mathrm{w},\;\qquad \qquad 
 e^{2\phi}=g_{s}^{2}H_{2}^{1/2}H_{4}^{-1/2},\label{C.2}\\
 & H_{2}=1+\frac{\Q_{2}}{r^{2}},\qquad H_{4}=1+\frac{\Q_{4}}{r^{2}},\qquad\ \ \ \Q_{2}=\frac{g_{s}\ls^2\, Q_{2}}{\mathrm{v}},\;\;\qquad \Q_{4}=\frac{g_{s}\ls^2 \, Q_{4}}{\mathrm{u}},\qquad\mathrm{v}=\mathrm{v}_{1}\mathrm{v}_{2}\mathrm{v}_{3} \ .
\label{C.3}
\end{align}
Here     $\Q_2$ and $\Q_4\equiv \Q_5$   are the same as in \rf{2.4} and we follow the notation introduced in \rf{2555}  and \rf{a222}, \rf{a00}, i.e. 
\begin{align}
&\mathrm{y}_{i}\equiv \mathrm{y}_{i}+2\pi\ls\mathrm{v}_{i},
 \qquad \mathrm{w}\equiv  \mathrm{w}+2\pi\ls\mathrm{u}, \qquad  \mathrm{v}_{i}=g_{s}^{\ftr}v_{i},\qquad 
 \mathrm{u}=g_{s}^{\ftr}u\ .  \label{C.4}
\end{align}
Taking the 
near-horizon limit   and rescaling $x_0$ and $x_1$ we get the  $\AdS_3\times S^{3}\times T^{3}\times S^{1}$
 type IIA   solution\foot{\la{f1}Note that the 
volume of the 4-torus $(\y_i,\w)$  in the near-horizon limit given by 
$\vol(T^4) =(2\pi)^4 \ls^4  { \Q_2\ov \Q_4}  \vv \uu=  (2\pi)^4 \ls^4  { Q_2\ov Q_4}  \uu^2$
does not depend on the  original 3-torus volume  parameter $\v$. 
}
\begin{align}
ds_{10A}^{2} & = \sqrt{\Q_{2}\Q_{4}}\big(ds_{\AdS_3}^{2}+ds^2_{S^3}\big)+\sqrt{\frac{\Q_{2}}{\Q_{4}}}d\mathrm y_{i}^{2}+\sqrt{\frac{\Q_{4}}{\Q_{2}}}d\mathrm w^{2},\qquad \qquad e^{2{\phi}}= \g^2=g_{s}^{2}\sqrt{\frac{\Q_{2}}{\Q_{4}}},\label{C.5}\\
\FF_{4} & =2g_{s}^{-1}\Q_{4}\big(i\, \text{vol}_{AdS_3}+\text{vol}_{S^{3}}\big)\wedge d\mathrm w.\label{C.6}
\end{align}
The  corresponding string coupling $\g$ and dimensionless tension
$\mathrm{T}$ are thus   as in \rf{155},\rf{55}\foot{Note that if  one    formally  
chooses $\Q_2=\Q_4$, i.e.  $\mathrm{v}^{-1} Q_2 =\uu^{-1} Q_{4}$ so  that
$\mathrm{{g}_{s}}=g_{s}$, then  $\mathrm{T}={}{g_{s}\uu^{-1}Q_{4}}$. }
\begin{align}
 &  \g^2 
 = \gs^2    \sqrt{ \uu\ov \mathrm v } \sqrt{  {Q_2\ov  Q_4}} , 
\qquad \ \  \qquad \T =\frac{L_s^{2}}{2\pi \ls^2}
 =\frac{\sqrt{\Q_{2}\Q_{4}}}{2\pi \ls^2}
 ={}\frac{g_{s}}{2\pi \sqrt{\mathrm{u}\mathrm{v}}}\sqrt{Q_{2}Q_{4}}\ .\label{C.7}
\end{align}
Similarly, the   D1-D5 solution of type IIB supergravity may be   represented as  
\begin{align}
 & d{s}_{10B}^{2}=H_{1}^{-1/2}H_{5}^{-1/2}\big(dx_{0}^{2}+dx_{1}^{2}\big)+H_{1}^{1/2}H_{5}^{1/2}\big(dr^{2}+r^{2}ds^2_{S^3}\big) +H_{1}^{1/2}H_{5}^{-1/2}d\y_{i}^{2}+H_{1}^{1/2}H_{5}^{-1/2}d\mathrm w'^{2}, \label{C.8}\\
 & {\FF}_{3}=g'^{-1}_{s}\big(i\, dH_{1}^{-1}\wedge dx_{0}\wedge dx_{1}-*_{4}dH_{5}\big),\qquad\qquad  
 e^{2{\phi'}}=\gs'^{2}H_{1}H_{5}^{-1}\ , 
   \label{C.9}\\
  & H_{1}=1+\frac{\Q_{1}}{r^{2}},\qquad\ \ \  H_{5}=1+\frac{\Q_{5}}{r^{2}},\qquad\ \
   \ \Q_{1}=\frac{g'_{s}\ls^2Q_{1}}{\hat \vv'},\qquad\ \ \ \Q_{5}=g'_{s}\ls^2Q_{5}\ , 
 \label{C.10}
\end{align}
where $\gs'$ is the  type IIB  asymptotic  string  coupling   and 
\begin{align}
&\mathrm y_{i}\equiv \mathrm y_{i}+2\pi\ls{\mathrm v}_{i},\qquad 
\ \ \w'\equiv \w'+{2\pi}\ls{\mathrm{u}'},\qquad\qquad  \mathrm v=\mathrm v_1 \mathrm v_2 \mathrm v_3,\qquad \ 
\hat \vv'= \vv \uu'  
  \ . \label{C.11}
\end{align}
The D2-D4  (\ref{C.1})  and  D1-D5
(\ref{C.8})  solutions are related by T-duality along the $\ww\to \ww'  $ direction, with the radii    measured  in units of $\ls$ 
 being related by   $\uu'=\uu^{-1}$  and the dilatons (and thus also the  asymptotic string couplings) 
   transforming as  $ \uu\,  e^{-2\phi}  = \uu'  e^{-2\p'}$  (and the RR potentials  related  according to the standard rule, see, e.g., 
 \ci{Bergshoeff:1995as}), i.e. 
\be 
\uu'=\uu^{-1} \ , \qquad \qquad \hat  \vv'= \vv \uu^{-1} \ , \qquad \ \ \   \gs'= \uu^{-1} \gs \ .  \la{C.100}
\ee
The harmonic functions   related by   $H_2=H_1, \ H_4=H_5$.

The near-horizon limit of \rf{C.8},\rf{C.9}   gives  the   $\AdS_3\times S^{3}\times T^{3}\times S^{1}$  IIB  solution
supported by RR flux\foot{\la{f2}Like in the above D2-D4 case  the 
volume of the 4-torus $(\y_i,\w)$  in the near-horizon limit  takes the form 
$\vol(T^4) =(2\pi)^4 \ls^4  { \Q_1\ov \Q_5} \hat  \vv'=  (2\pi)^4 \ls^4  { Q_1\ov Q_5} $ and thus  is fixed, i.e. 
does not depend on the original 4-torus volume  parameter $\hat \vv'$  \ci{Maldacena:1997re,Aharony:1999ti}. 
}
\begin{align}
 & d{s}{}_{10B}^{2}=\sqrt{\Q_{1}\Q_{5}}\big(ds_{\AdS_3}^{2}+ds^2_{S^3}\big)+\sqrt{\frac{\Q_{1}}{\Q_{5}}}(d\y_{i}^{2}+d\w'^{2})\ ,\qquad\qquad  e^{2{\phi}'}= \gs'^2 \frac{\Q_{1}}{\Q_{5}}\ , \label{C.12}\\
 & {\FF}_{3}=2\gs'^{-1}\Q_{5}(i\, \text{vol}_{AdS_3}+\text{vol}_{S^{3}}) \ . \label{C.13}
\end{align}
The corresponding 
string coupling $\g'$ and dimensionless tension ${\T'}$  here are as
 in\rf{136},\rf{505}\foot{If  one formally  sets  $\Q_1=\Q_5$, i.e. 
  $\hat \vv'^{-1} Q_5=Q_{1}$   and   $g_{6}\equiv {1\ov \sqrt{\hat \vv'}}  \gs'$   
  (as, e.g.,   in \cite{Aharony:2024fid})
then  $\g'=g'_{s}$ and   $\T' = \gs' Q_5= g_{6}\sqrt{Q_{1}Q_{5}}.$ }
\begin{align}
 & \g'
 =\frac{g'_{s}}{\sqrt{\hat \vv'}}\sqrt{\frac{Q_{1}}{Q_{5}}}\ ,\qquad \qquad
 \mathrm{T}'= \frac{\sqrt{\Q_{1}\Q_{5}}}{2\pi \ls^2}
 =\frac{g_{s}'}{2\pi \sqrt{\hat \vv'}}\sqrt{{Q_{1}Q_{5}}}\ . \label{C.14}
\end{align}

\section{Supersymmetry of $\AdS_2 \times S^1$   M2 brane  solution
\label{apa}}

The bosonic   and quadratic   fermionic  parts  part of the  M2-brane action in 11d  supergravity 
background   may be written as (we use Euclidean signature)  \ci{Bergshoeff:1987cm}
\begin{align}
    &S_{\rm B}=S_{\rm V}+S_{\rm WZ},\qquad \qquad 
    S_{\rm V}=T_{2}\int d^{3}\sigma\sqrt{h},\qquad\qquad  h_{mn}=\partial_{m}X^{M}\partial_{n}X^{N}G_{MN}(X)\label{A.1}\ , \\
 &\la{A11} 
  S_{\rm WZ}=-\tfrac{i}{3!}T_{2} \int d^{3}\sigma\ \varepsilon^{mnp}C_{MNP}(X)\partial_{m}X^{M}\partial_{n}X^{N}\partial_{p}X^{P}\ ,  \\
   &S_{\rm F}=iT_2 \int d^3\s \Big[\sqrt{h}h^{mn}\partial_mX^M\bar \theta \Gamma_M \hat D_n \theta-\tfrac i2 \epsilon^{mnk}\partial_m X^M \partial_n X^N \bar \theta \Gamma_{MN}\hat D_k \theta \Big],\qquad \ \ \  \Gamma_M=E^A_M\Gamma_A\ , \label{A.5}\\
     &\hat D_m=\partial_m X^M \hat D_M, \qquad  \hat D_M=D_M-\tfrac 1{288}(\Gamma^{PQRS}\,_M+8\Gamma^{PQR}\delta^S_M)F_{PQRS},\qquad D_M=\partial_M+\tfrac 14 \Gamma_{AB}\omega_M^{AB} . \no 
     \end{align}
To preserve   supersymmetry  an   M2 brane   solution  $X^M(\s)$ should  satisfy 
\begin{equation}
    \Gamma_{\ka}\, \epsilon=\epsilon,\qquad\qquad \Gamma_{\ka}=\frac{i}{3!\sqrt{h}}\varepsilon^{mnp}\partial_{m}X^{M}\partial_{n}X^{N}\partial_{p}X^{P}\Gamma_{MNP}\ ,\label{A.4}
\end{equation}
where  $\epsilon$ is a Killing spinor of the background, 
\be
\delta \psi_M= \hat D_M\epsilon \equiv ( \del_M - \Omega_M) \epsilon=0 \   .\la{D.1} 
\ee
To show that  the  $\mathrm{AdS}_2\times S^1$ M2 brane  solution  preserves  half of   supersymmetry
let us  first 
   find explicitly the  solution of \rf{D.1} for the Killing spinor 
    in  the case of \adsts\ background in \rf{2.5}
    with \AdS$_3$ and $S^3$ parametrized 
  as (cf. \rf{3.7} 
\begin{equation}
    ds_{\AdS_3}^{2}= \cosh^2 \zeta \,(\cosh^2 \xi\, d\eta^2+d\xi^2) + d\zeta^2,\qquad\qquad 
      ds_{S^3}^{2}=d\rho^2 + \sin^2\rho\, d\alpha ^2 +\cos^2\rho\,  d\beta^2\ . \label{D.3}
\end{equation}
We shall   label  the  11d coordinates as  $X^{M}=(X^0, X^1, ..., X^{10})= (\eta,\xi,\zeta,\rho,\alpha,\beta,\mathrm y^{1},\mathrm y^{2},\mathrm y^{3},\mathrm y^{4},\mathrm w)$. The  
  underlined    indices  will stand for tangent space components.  Then \rf{D.1}   may be  written  as 
\begin{align}
 &\te  \delta\psi_{\eta}=\partial_{\eta}\epsilon+\frac 1 2\sinh \xi\,\Gamma^{\underline{01}}+\frac{1}{2}\cosh \xi\sinh \zeta\,\Gamma^{\underline{02}}\epsilon-\frac{i}{6}\cosh \zeta\cosh \xi\,\Gamma^{\underline{12\,10}}\big(\Gamma_*-2\big)\epsilon=0\ ,\label{D.4}\\
 &\te  \delta\psi_{\xi}=\partial_{\xi}\epsilon+\frac{1}{2}\sinh \zeta\,\Gamma^{\underline{12}}\epsilon-\frac{i}{6}\cosh \zeta\,\Gamma^{\underline{02\,10}}\big(\Gamma_*-2\big)\epsilon=0\ , \qquad\qquad  \ \  \Gamma_*\equiv i\Gamma_{\underline{012345}}\ ,   \label{D.5}\\
 &\te  \delta\psi_{\zeta}=\partial_{\zeta}\epsilon-\frac{i}{6}\Gamma^{\underline{01\,10}}\big(\Gamma_*-2\big)\epsilon=0,\qquad\qquad  \delta\psi_{\rho}=\partial_{\rho}\epsilon-\frac{1}{6}\Gamma^{\underline{45\,10}}\big(\Gamma_*-2\big)\epsilon=0\ , \label{D.7}\\
 &\te  \delta\psi_{\alpha}=\partial_{\alpha}\epsilon-\frac{1}{2}\cos\rho\,\Gamma^{\underline{34}}\epsilon+\frac{1}{6}\sin\rho\,\Gamma^{\underline{35\,10}}\big(\Gamma_*-2\big)\epsilon=0\ , \label{D.8}\\
 &\te  \delta\psi_{\beta}=\partial_{\beta}\epsilon+\frac{1}{2}\sin\rho\,\Gamma^{\underline{35}}\epsilon-\frac{1}{6}\cos\rho\,\Gamma^{\underline{34\,10}}\big(\Gamma_*-2\big)\epsilon=0\ , \label{D.9}\\
 &\te  \delta\psi_{\mathrm y_p}=\partial_{\mathrm y_p}\epsilon-\frac{R}{6L}\Gamma_{\underline{\mathrm y_p}}\Gamma^{\underline{345\,10}}\big(\Gamma_*+1\big)\epsilon=0,\qquad \qquad \delta\psi_{\mathrm w}=\partial_{\mathrm w}\epsilon-\frac{1}{6LR^{2}}\Gamma^{\underline{345}}\big(\Gamma_*+1\big)\epsilon=0\ . \label{D.11}
\end{align}
We shall   consider  the  Killing spinor  to be 
 constant in the $T^{5}$ directions $(\mathrm y_p, \mathrm w)$ by 
 choosing a projection 
 \begin{equation}
\label{D.12}
 \Gamma_*\, \epsilon =- \epsilon \ ,    
\end{equation}
so that  the equations in \rf{D.11} are satisfied. It remains to solve 
 the   6 equations in  \rf{D.4}--\rf{D.9}.  As the   two   components of $\Omega_M$ in \rf{D.1} 
given by   ${\Omega}_{\zeta}=-\frac{i}{2}\Gamma^{\underline{01\,10}}$ and ${\Omega}_{\rho}=-\frac{1}{2}\Gamma^{\underline{45\,10}}$ are constant,  we get 
\begin{equation}
    \epsilon=\, {\exp}\big[{\Omega}_{\zeta}\zeta\big]\, {\exp}\big[{\Omega}_{\rho}\rho\big]\,\chi\ ,  \label{D.17}
\end{equation}
where $\chi$  depends on  the other four coordinates. 
The remaining equations  written in terms of $\chi$  are 
\begin{align}
&\partial_{\eta}\chi=  \hat{\Omega}_{\eta}\chi\ , 
\qquad  \partial_{\xi}\chi= \hat{\Omega}_{\xi}\chi\ , 
\qquad \partial_{\alpha}\chi= \hat{\Omega}_{\alpha}\chi\ , 
\qquad \partial_{\beta}\chi=  \hat{\Omega}_{\beta}\chi\ , \label{D.18}
\\
&\te \hat{\Omega}_{\xi}=e^{-{\Omega}_{\zeta}\zeta}\, \Omega_{\xi}(\zeta) \, e^{{\Omega}_{\zeta}\zeta}=\frac i2\Gamma^{\underline{02\,10}}\ ,\qquad 
\qquad\hat{\Omega}_{\eta}=e^{-\hat{\Omega}_{\xi}\xi}\ e^{-{\Omega}_{\zeta}\zeta}\, \Omega_{\eta}(\zeta)\, e^{{\Omega}_{\zeta}\zeta}\, e^{\hat{\Omega}_{\xi}\xi}=-\frac{i}{2}\Gamma^{\underline{12\,10}}\ ,\label{D.19}\\
&\te \hat{\Omega}_{\beta}=e^{-{\Omega}_{\rho}\rho}\, \Omega_{\beta}(\rho)\, e^{{\Omega}_{\rho}\rho}=-\frac{1}{2}\Gamma^{\underline{34\,10}}\ ,\qquad
\qquad \hat{\Omega}_{\alpha}=e^{-\hat{\Omega}_{\beta}\beta}\, e^{-{\Omega}_{\rho}\rho}\, \Omega_{\alpha}(\rho)\, e^{{\Omega}_{\rho}\rho}\, e^{\hat{\Omega}_{\beta}\beta}=\frac{1}{2}\Gamma^{\underline{34}}\label{D.20}\ ,
\end{align}
  and thus  are solved by 
$\chi=\, {\exp}\big[\hat{\Omega}_{\xi}\xi\big]\, {\exp}\big[\hat{\Omega}_{\eta}\eta\big]\, {\exp}\big[\hat{\Omega}_{\beta}\beta\big]\, {\exp}\big[\hat{\Omega}_{\alpha}\alpha\big]\epsilon_{0}$. 
The final  expression  for the Killing  spinor is then 
\begin{equation}
\label{D.25}
\epsilon=\, {\exp}\big[{\Omega}_{\zeta}\zeta\big]\, {\exp}\big[\hat{\Omega}_{\xi}\xi\big]\, {\exp}\big[\hat{\Omega}_{\eta}\eta\big]\, {\exp}\big[{\Omega}_{\rho}\rho\big]\, {\exp}\big[\hat{\Omega}_{\beta}\beta\big]\, {\exp}\big[\hat{\Omega}_{\alpha}\alpha\big]\, \epsilon_{0},\qquad \ \ \ \ \ \ \ \Gamma_*\, \epsilon_0=-\epsilon_{0}\ , 
\end{equation}
where  we imposed  the projection on $\epsilon_0$  to satisfy the condition \rf{D.12}. 
This demonstrates that 
the $\ads_3 \times S^3 \times T^5$  background has 16 superisometries  parametrised by $\epsilon_0$, 
 i.e.  it preserves $16$ supercharges. 


The $\mathrm{AdS}_2\times S^1$ M2 brane   solution 
 corresponds to $\eta$ and $\xi$  in \rf{D.3} identified with $\s^1$ and $\s^2$ world-volume coordinates,  
 $\varphi$ in \rf{277}  with $\sigma^3$  and 
\begin{equation}\la{d20}
\zeta=0,\qquad\qquad  \rho=0,\qquad\qquad  \beta=0,\qquad \qquad \alpha=\alpha_0 \ . 
\end{equation}
The   condition \rf{A.4} that it  preserves supersymmetry may   then  be written as 
\begin{equation}
\Gamma_\ka\,  \epsilon
= i\Gamma_{\underline{019}}\, \epsilon
=\epsilon \ . \la{D21} 
\end{equation}
Using \rf{D.25} one can check  that  \rf{D21}   is satisfied  provided $\epsilon_0$ is subject to the same 
 projection  \rf{D21}   and also the one in \rf{D.25}, i.e. 
\begin{equation}
\Gamma_\ka\,  \epsilon_0=\epsilon_0\ ,
\qquad\qquad 
\Gamma_*\, \epsilon_0=-\epsilon_0\ ,
\qquad\qquad 
\big[\Gamma_\ka,\Gamma_*\big]=0.
\end{equation}
As a result, $\epsilon_0$ has 8 independent components, i.e.
the  $\mathrm{AdS}_2\times S^1$ M2 brane   solution  
 preserves half of the  supersymmetry of the \adsts\ background.

\section{\AdS$_2$ supersymmetry   and mass  spectrum  
\la{apd}}

The
$\mathrm{AdS}_3 \times S^3 \times T^5$ solution of  11d 
supergravity  preserves $16$ real supercharges,
corresponding to $\mathcal{N}=(4,4)$ supersymmetry  from  the dual  2d 
 CFT  point of view (see, e.g., \ci{deBoer:1998kjm}). 
The bosonic isometry group  has the  algebra  $\mathfrak{so}(2,2) \oplus \mathfrak{so}(4) \oplus \mathfrak{u}(1)^5 $.
Ignoring the last  torus symmetry factor  this  background  has global  superisometry  algebra 
 $\mathfrak{psu}(1,1|2)_L\times \mathfrak{psu}(1,1|2)_R $.\foot{This is  
 the global part of the
``small'' $\mathcal{N}=(4,4)$ superconformal symmetry in two dimensions.}

The presence of an $\ads_2\times S^1$  M2 brane defect breaks it 
 to   $\mathfrak{psu}(1,1|2)\times \mathfrak{su}(2)_A$ (see, e.g., \ci{Bliard:2024bcz}). 
  The bosonic part of  $\mathfrak{psu}(1,1|2)$ is $\mathfrak{sl}(2,R)\times \mathfrak{su}(2)_R$, where $\mathfrak{sl}(2,R)$  is the
  global   1d conformal  algebra 
  (with generators $D,P,K$)  corresponding to \AdS$_2$ 
   and $\mathfrak{su}(2)_R$ is the R-symmetry (with generators $R^\aa_\bb$, \   $\aa, \bb=1,2$).
   $\mathfrak{su}(2)_A$ acts as an automorphism with generators $A^{\dot \aa}_{\dot \bb}$.
Omitting  the commutators involving $R^\aa_\bb$ and $A^{\dot \aa}_{\dot \bb}$  we have 
\begin{align}
&[D,P]=P,\qquad [D,K]=-K,\qquad [P,K]=-2D,\qquad  [D,Q_{\aa \dot \aa }]=\tfrac 12 Q_{\aa \dot \aa},\qquad [D,S_{\aa \dot \aa }]=-\tfrac 12 S_{\aa \dot\aa },\\
&[P,S_{\aa \dot\aa }]=-Q_{\aa \dot \aa},\qquad [K,Q_{\aa \dot \aa}]=S_{\aa \dot \aa},\qquad  \{Q_{\aa \dot \aa},Q_{\bb \dot \bb}\}=-\epsilon_{\aa \bb}\epsilon_{\dot \aa \dot \bb}P\ ,\\
&\{Q_{\aa \dot \aa},S_{\bb \dot \bb}\}=-\epsilon_{\aa \bb}\epsilon_{\dot \aa \dot \bb}D-\epsilon_{\dot \aa \dot \bb}R_{\aa \bb}, \qquad \qquad \{S_{\aa \dot \aa},S_{\bb \dot \bb}\}=-\epsilon_{\aa \bb}\epsilon_{\dot \aa \dot \bb}K\ . \la{d3}
\end{align}
The  four  2d spinor  generators  $Q_{\aa \dot \aa}$ transform as  $(\mathbf 2,\mathbf 2)$ under $SU(2)_R\times SU(2)_A$.
The states  in 
supermultiplets can be labelled by their  1d scaling dimension $\Delta$, i.e. $D|\Delta\rangle=\Delta |\Delta \rangle$, 
as well as  quantum numbers corresponding to $R^\aa_\bb$ and $A^{\dot \aa}_{\dot \bb}$ that we shall omit
(see, e.g., \ci{deBoer:1998kjm}). 

The long  $\N=(4,4)$  supermultiplet of $\mathfrak{psu}(1,1|2)$
 is obtained by acting  on   a given  state $|\Delta\rangle$ with $Q_{\aa \dot \aa}$
\begin{equation}\te 
 1\times |\Delta \rangle,\qquad 4\times |\Delta +\frac 12\rangle,\qquad 6\times  |\Delta +1\rangle,\qquad4\times  |\Delta +\frac 32\rangle,\qquad 1\times |\Delta +2 \rangle.\la{d4}
\end{equation}
The corresponding  masses   of scalars and fermions propagating in \AdS$_2$   are related to the values of $\Delta$ as 
$m^2_{\rm B}= \Delta(\Delta-1), \ |m_{\rm F}|=  \Delta -\ha $, i.e. 
\begin{equation}
\text{scalar}: \ \ \Delta=\tfrac 12 \Big(1+\sqrt{1+4m_{\rm B}^2}\Big),\qquad\qquad  \text{spinor}: \ \ \Delta =\tfrac 12\Big(1+2|m_{\rm F}|\Big) \ . \la{d5} 
\end{equation}
The resulting spectrum of 8+8  fields in \AdS$_2$  is then 
\begin{align}
&{\text{scalars} }\ \  \;m_{\rm B}^2:\,\qquad \Delta(\Delta-1)\; \text{(1 state)},\qquad \;(\Delta+1)\Delta\; \text{(6  states)}, \qquad \;(\Delta+1)(\Delta+2)\; \text{(1 state)}\ , \\
&\text{spinors }\ \ \;|m_{\rm F}|:\;\qquad \Delta\;\text{(4 states)},\qquad \;\qquad\;\;\;\qquad\Delta+1\;\text{(4 states)}\ . \la{d7}
\end{align}
This  agrees with the  quadratic fluctuation  spectrum of the M2 brane 
 represented in terms of the $S^1$  Fourier  modes 
 propagating on \AdS$_2$ if  (see \rf{4.42})
 \be  \Delta_n =\tfrac 12(m_n-1)\ ,\qquad \qquad   \;m_n=\sqrt{1+4\k^2 n^2} \ , \qquad n\not=0 \ .  \la{dd8}  \ee
Note that the $n=0$ fluctuations fit into  short multiplets  corresponding to 
the displacement and tilt multiplets from the boundary theory   point of view.  


Let us now  comment on the  supermultiplet structure of quadratic fluctuations 
 in the case   of  AdS$_2\times S^1$   M2  brane in 
 AdS$_4 \times S^7/\mathbb{Z}_k$  considered  in \ci{Sakaguchi:2010dg,Giombi:2023vzu}.
 Here for $k >2$   the relevant  supersymmetry  algebra is expected to be $\mathfrak{su}(1,1|3)$
 (with the bosonic part  $\mathfrak{su}(1,1)\oplus\mathfrak{su}(3)\oplus \mathfrak{u}(1)$)
which  is   the subalgebra of $\mathfrak{osp}(6|4)$    symmetry of $\ads_4\times \mathbb{CP}^3$  
 that is preserved by a 1d defect   (see, e.g., \ci{Bianchi:2017ozk,Bianchi:2020hsz,Gorini:2022jws}). 
 There is also an extra   $ \mathfrak{u}(1)_b $  symmetry   associated to $S^1$ of AdS$_2\times S^1$  M2  brane in 
 AdS$_4 \times S^7/\mathbb{Z}_k$ 
 corresponding to the KK   charge $n$.\foot{
 For $k=1,2$ the symmetry of the background enhances from $\mathfrak{osp}(6|4)$ to $\mathfrak{osp}(8|4)$ and 
   the maximal 1d $\ha$-BPS subalgebra preserved by the defect  is $\mathfrak{su}(1,1|4)\oplus \mathfrak{u}(1)_b$.} 
 
 The  spectrum of  towers of  quadratic fluctuations in AdS$_2$ 
   that leads  to the 1-loop   partition function in \rf{9},\rf{10} 
 here is given by  ($|kn| > 2$,  $n \in \mathbb{Z}$):  
  (i)   complex scalars $\eta_n$ (two real  for each $n$) with 
$
m^2 = \frac{1}{4}(kn-2)(kn-4)$;    3 sets of complex scalars $\zeta^a_n$ ($a=1,2,3$) with 
$
m^2 = \frac{1}{4}kn(kn+2)$;
(ii) 8  two-component spinor fields  $\vartheta^A_n$   with 
$
m = \frac{1}{2}kn \pm 1 $ $(\mbox{3+3 modes})$   and 
$m = \frac{1}{2}kn $  $(\mbox{2 modes})$.  
These   can be organized into $\N=1$ multiplets  with $m_{\rm B}^2 = m_{\rm F} (m_{\rm F}-1)   $
 \ci{Giombi:2023vzu}.

The $n=0$ modes    correspond to the   short multiplet of  $\mathfrak{su}(1,1|3)$,  while  the $n\not=0$ modes  are
a priori  expected 
to   belong  to    long multiplets. 
The long multiplet of $\mathfrak{su}(1,1|3)$  contains 32+32  states 
obtained by  acting with all 6 supercharges ($Q_a$ and $\bar Q^a$,   $a=1,2,3$).\foot{The full long multiplet is given by 
$
[\Delta;1]_\rB
\oplus
[\Delta+\tfrac12;\,3\oplus\bar{3}]_\rF
\oplus
[\Delta+1;\,8\oplus1\oplus 3\oplus\bar{3}]_\rB
$ $
\oplus\;
[\Delta+\tfrac32;\,2\times 1\oplus3\oplus\bar{3}\oplus6\oplus\bar{6}]_\rF
\oplus
[\Delta+2;\,8\oplus1\oplus3\oplus\bar{3}]_\rB
$ $
\oplus\;
[\Delta+\tfrac52;\,3\oplus\bar{3}]_\rF
\oplus
[\Delta+3;1]_\rB.
$
To get the full long multiplet one would  need to combine states from different   $n$-levels  as at each level we have only 8+8   states. } 
To exhibit the extended   supersymmetry structure of the massive fluctuation   mode spectrum 
we may organize them  for each $n$ 
into    ``chiral''  sectors $(\zeta_n^a, \eta_n) $   and $(\bar\zeta_n^a, \bar\eta_n)$  combined with the corresponding fermions. 
These   will form  conjugate 
  ``shortened''   supermultiplets  corresponding to the action of $Q_a$ and $\bar Q^a$, respectively, i.e.  
obtained   by  imposing  $ \bar Q^a|\Delta\rangle =0$  or  $
Q_a|\Delta\rangle =0$.  The    
 shortening condition     (see (C.26) and (C.30) in \cite{Bianchi:2017ozk})  is 
$
\Delta= \mp\frac 13j_0,\ \ j_1=j_2=0$
($j_0,j_1,j_2$ are Dynkin labels of  Cartan generators of $\mathfrak u(1)$ and $\mathfrak{ su}(3)$).

The shortened multiplet  contains 4+4  states 
$
1\times |\Delta\rangle ,\ \  3\times |\Delta+\tfrac 12\rangle ,\ \ 
3\times |\Delta+1\rangle,\quad 1\times |\Delta+\tfrac 32\rangle
$. 
 If    the lowest $\Delta$ state   is  bosonic  (so we have $1_\rB \to 3_\rF \to \bar 3_\rB \to 1_\rF$) 
  then the corresponding masses   are 
$
m_\rB^2:\  1 \times   \Delta(\Delta-1), \  3\times  (\Delta+1)\Delta $  and 
$m_\rF:\   3\times  \Delta , \  1\times  (\Delta+1)$.
The above quadratic   fluctuation spectrum at given $n$ can be 
 organized into the sum of  two  such shortened  multiplets (corresponding to 
   $Q_a$ and $\bar Q^a$  and  related by the conjugation  $n\to -n$)
 with the  lowest   $\Delta> \ha $  given by  $\Delta_{1,n}$ or $\Delta_{2,n}$  (cf. \rf{dd8}) 
 \be
\Delta_{1,n} =\tfrac 12(kn+1),\qquad \qquad \Delta_{2,n}=\tfrac 12(kn-2),\qquad \qquad kn\geq3 \ . 
\label{d22}
\ee
Computing the  value of the 1-loop   contribution corresponding   to the above   shortened  multiplet 
 using \rf{3.48}--\rf{3.57}   one finds as in \rf{3344}   that all transcendental parts cancel out  and one  gets   simply:
 \ 
 $\zeta'_{\text{tot}, \Delta}  (0) = -\log\Delta $  (if $ |\Delta\rangle$    is  a  bosonic state) and 
 $\zeta'_{\text{tot}, \Delta}  (0) = -{\rm log}(\Delta + \tfrac{1}{2})$  (if $ |\Delta\rangle$ is a    fermionic state).
Considering the  contribution to the effective action in \rf{3.48}  from all  the states at levels $n$ and $-n$ 
(that form  two multiplets with $\Delta_{1,n}$ and two multiplets with $\Delta_{2,n}$)  and using \rf{d22}   we get ($k>2$)
\be\la{d23}
\zeta'_{\text{tot},n}+\zeta'_{\text{tot},-n}=-2\log ( \Delta_{1,n} + \ha)- 2 \log \Delta_{2,n}=-2\log \left[\tfrac 14(n^2k^2-4)\right]\ ,  
\ee
in agreement with \rf{011}.  We conclude that like in the AdS$_3 \times S^3$   case discussed in section 4 
 here  the  non-trivial   cancellations in the 1-loop effective action in \rf{9},\rf{10}  leading to \rf{011}   can be understood as a consequence  of  the corresponding  extended  AdS$_2$ supersymmetry.

\newpage
\section{Mixed flux  case   \la{ape}}

D1-D5    solution  is  related  by  S-duality   to NS1-NS5    one  and more generally 
to   $(\pp,\qq)$ family of $1\ov 4$-supersymmetric   type IIB  solutions 
 supported  by a combination of  RR   and NSNS   fluxes.
 This background  is  T-dual to  a generalization of  type IIA  D2-D4  solution 
  supported  by a combination of   fluxes. 
  
  From the 11d point of  view, the   S-duality of type IIB 
   theory  on $S^1$  is a  modular transformation on a 2-torus \ci{Bergshoeff:1995as,Schwarz:1995dk,Aspinwall:1995fw,Schwarz:1995jq,Witten:1995im}.
    Indeed, as we shall discuss    below,  
    the  $(\pp,\qq)$  generalization  of the  D2-D4 solution 
    can be obtained   from  the  M2-M5  solution  in \rf{2.1}--\rf{2.4} 
    upon dimensional reduction along   11th  direction 
     which is   a linear combination of 
    the  isometric   circular  directions $\y_4$ and $\ww$  
     \ci{Russo:1996if,Tseytlin:1997cs}. 
    
    The  near-horizon limit of the above type  IIB solution is \adst\   supported by   mixed  3-form   flux, its type IIA 
    T-dual  is supported by $F_4$ and $\H_3$ flux    and   their  11d  counterpart is   the same \adsts\  metric 
    supported by $F_4$ as 
     in \rf{2.5},\rf{277}  but  with the  $S^1$   direction $w$ replaced by  its linear  combination with $y_4$.
     It is then straightforward  to  find  the corresponding 
      generalization  of the  AdS$_2\times S^1$ M2  brane solution 
     discussed  in section 3.  Its  type IIA  string counterpart will be   the  AdS$_2$  minimal surface representing WL 
       in the  ``mixed flux'' background.

     \subsection{(D1-D5)$_{\pp,\qq}$    background   and its type   IIA  counterpart}
     
     Let us start with the ($\pp,\qq)$  family  of solutions 
     that includes   the NS1-NS5  \ci{Tseytlin:1996as}  and D1-D5 \ci{Callan:1996dv} solutions
        of type IIB  supergravity,  
     i.e.   which is  a    generalization of  \rf{C.8}--\rf{C.10}. 
     It is obtained by  applying a  special  $SO(2) \subset SL(2,\mathbb R)$ transformation  \ci{Schwarz:1995dk} 
       which is a symmetry of the type IIB  equations of motion\foot{The type IIB       action  which is the analog of \rf{B.7} 
       has the structure (prime  on $\gs$ and $\tilde \phi$ is used to indicate type IIB  quantities): 
      \ \  
     $
  S_{10B}=\frac{1}{2\kappa^{2}_0  g_s'^2 } \int d^{10}x\sqrt{-G}\big[ e^{-2\tphi'}\big(R+4(\partial\tphi')^{2}-\tfrac{1}{2}\big|\H_{3}\big|^{2}\big) -\tfrac{1}{2} \big|{F}_{3}- C\,  \H_3\big|^2+...\big]  $, where  $F_3= g'_s \FF_3$. }
\al{
& F_3\  \rightarrow\  \cos \gamma\, F_3 -\sin\gamma\, \H_3\ ,\qquad\qquad 
  \H_3\ \rightarrow\  \cos\gamma\,  \H_3+\sin\gamma \, F_3\ ,\qquad \ \ \ \    F_3= g'_s \FF_3\  ,\la{ee2} \\
&\tau\equiv  C+ie^{-\tilde \phi'} \ \  \rightarrow  \ \
\frac{ \cos \gamma \,  \tau - \sin \gamma }{ \sin \gamma \, \tau  + \cos \gamma} =
 \frac{\mathrm p\,\tau-
g_s'^{-1}\mathrm q}{
g_s'^{-1}\qq\,\tau + \pp  } \ ,\qquad  \cos\gamma\equiv \frac {\mathrm p}{\sqrt{
\pp^2    + g_s'^{-2}\mathrm q^2}}\ ,  \quad  \mathrm p,\mathrm q\in \mathbb Z\ 
. \la{ee3} 
}
Here   we  follow the same  notation as in  \rf{C.8}--\rf{C.10}:  $g_s'$ is the asymptotic coupling of type IIB string 
(which is   not transformed as we are looking for a more general solution in the same theory)  and $\tilde \phi'$ is the ``non-constant'' part of the type IIB dilaton, i.e. $e^{\phi'}= g_s'  e^{ \tilde \phi'}$.
The resulting   generalization of the D1-D5 solution is then\foot{It   is equivalent to  the solution in  Eq. (2.15)  of \ci{Russo:1996if}   where 
$g_s'$ was set to 1. 
As in    \rf{C.8}    here  the  metric in  \rf{e4} is the string-frame one.}
\al{
& {ds}_{10B}^2=
H_{15}^{1/2}   \Big[ H_{1}^{-1} H_{5}^{-1/2}\big(dx_{0}^{2}+dx_{1}^{2}\big)+
H_{5}^{1/2}\big(dr^{2}+r^{2}ds^2_{S^3}\big)  + H_{5}^{-1/2} (d\y_{i}^{2}+ d\mathrm w'^{2}) \Big] \ , \la{e4}\\
&\FF_3=\cos\gamma  \,
g_s'^{-1} (i\, dH_1^{-1}\wedge dx_0\wedge dx_1-*_4 dH_5),\qquad\quad \H_3=\sin\gamma\,(i\, dH_1^{-1}\wedge dx_0\wedge dx_1-*_4 dH_5),\label{e6}\\
&C=\cos\gamma \sin\gamma (H_5-H_1)H_{15}^{-1},\qquad\ \ \  \qquad e^{2\phi'}=g_s'^2H_{15}^2H_1^{-1}H_5^{-1},
\la{e7}\\
& H_{1,5}=1+\frac {\bar  \Q_{1,5}}{r^2}\ , \qquad \ \  \bar  \Q_{1,5} \equiv 
\sqrt{ \mathrm p^2+g_s'^{-2}\mathrm q^2}\  \mathcal Q_{1,5}\ , \ \ 
\qquad \ \ 
H_{15}\equiv \cos^2\gamma\, H_1+\sin^2\gamma\, H_5 \ , \la{e5}
 }
where    $ \Q_{1}={g'_{s}\ls^2\hat \vv'^{-1} Q_{1}}$  and $ \Q_{5}=g'_{s}\ls^2Q_{5}$
are the same as in \rf{C.10}.

In general, this background  describes  a system with 4 sets of integer  charges:
\be \la{e567} Q_{1 \rm RR}= \pp Q_1, \qquad   Q_{1 \rm NS}= \qq Q_1, \qquad  Q_{5 \rm RR}= \pp Q_5, \qquad
  Q_{5 \rm NS}= \qq Q_5 \ . \ee
The original  D1-D5  solution   corresponds to  the   special case of  p=1, q=0  ($\gamma=0$)
 while the NS1-NS5 one to  p=0,  q=1  ($\gamma = {\pi\ov 2}$)
(in the latter   case   $\bar  \Q_{1} =  \ls^2 \hat {\rm v}'^{-1}  Q_1    , \ \   \bar  \Q_{5} =  \ls^2   Q_5$).

The  near-horizon limit  of this solution    is  represented   by 
the  $\ads_3\times S^3\times T^4$  supported by a  mixed  3-form flux (with RR scalar $C$ and dilaton being constant)\foot{We again rescale $x_0$ and $x_1$  as in AdS$_3$ in \rf{C.12}.}
\al{
&\qquad    {ds}_{10B}^2 =\td L_s^2\Big[ds^2_{\ads_3} +ds^2_{S^3}+{\bar \Q_5} ^{-1}(d\mathrm y_{i}^{2}+d\w'^{2})\Big],\qquad\la{ee7}
  \\
  \td L_s^2&=(\cos^2\gamma\, \bar\Q_1+\sin^2\gamma\, \bar\Q_5)^{1/2}\bar \Q_5^{1/2}=   
   (\pp^2\Q_1 \Q_5   +\gs'^{-2}  \qq^2 \Q_5^2)^{1/2} =    ( \pp^2  \hat \vv'^{-1}  \gs^2 Q_1 Q_5 + \qq^2 Q_5^2)^{1/2}\ls^2  \ ,\la{e97}\\
\mathrm F_3&=2\cos\gamma\,g_s'^{-1}\, \bar \Q_5\,  (i\, \text{vol}_{\ads_3}+\text{vol}_{S^3} ) 
 =   
   2\pp\, Q_5\ ( i\, \text{vol}_{\ads_3}+\text{vol}_{S^3}  ) \ls^2 \  ,\la{e002}\\
 \mathrm H_3&=2\sin\gamma\, \bar \Q_5\, (i\, \text{vol}_{\ads_3}+\text{vol}_{S^3} )
 =     2\qq\,  Q_5\,  (i\, \text{vol}_{\ads_3}+\text{vol}_{S^3} )  \ls^2 \  
   ,\la{e001} \\
C&=\frac{\cos\gamma \sin\gamma(\mathcal Q_5-\mathcal Q_1)}{\cos^2\gamma \mathcal Q_1+\sin^2\gamma \mathcal Q_5},
\qquad\ \ \qquad\  e^{2\phi'}=g_s'^2\frac{(\cos^2\gamma \mathcal Q_1+\sin^2\gamma \mathcal Q_5)^2}{\mathcal Q_1\mathcal Q_5}\ .\la{e9}
}
This reduces  to  \rf{C.12})     when $\pp=1, \qq=0$  ($\cos \gamma =1$)  and thus $\bar \Q_{1,5}=\Q_{1,5}$.

  Applying  the same  T-duality  transformation 
  along  the $\rm w$ direction  
    that related    type IIA  \rf{C.1}--\rf{C.3}  and type IIB \rf{C.8}--\rf{C.10} solutions 
     we find for the type IIA   counterpart of \rf{e4}--\rf{e5}
\al{
&\la{e10} ds_{10A}^{2}
=H_{24}^{1/2}\Big[
H_{2}^{-1}H_{4}^{-1/2}\big(dx^2_{0}+dx^2_{1}\big)
+H_{4}^{1/2}\big(dr^2+r^2 ds_{S^{3}}^{2}\big)
+H_4^{-1/2}d\y_i^2\Big]+H_{24}^{-1/2}H_4^{1/2}d\w^2\ , \\
&\la{e12} \FF_{4}=\cos\gamma \,g_s^{-1}(i\, dH_{2}^{-1}\wedge dx_{0}\wedge dx_{1}-*_{4}dH_{4})\wedge d\w,
\quad 
\H_3=\sin \gamma \,(i\, dH_{2}^{-1}\wedge dx_{0}\wedge dx_{1}-*_{4}dH_{4}), \\
&\la{e13} 
C_{\rm w} ={\cos \gamma \sin \gamma\, (H_4-H_2) H_{24}^{-1}},
\qquad \quad
e^{2\phi}=g_s^2H_2^{-1}H_4^{-1/2} H_{24}^{3/2},\\
&\la{e11}   H_{2,4}\equiv H_{1,5}=1+\frac{\bar {\mathcal Q}_{2,4}}{r^2},
\qquad \ \ \ 
\bar{\mathcal Q}_{2,4}\equiv \bar{\mathcal Q}_{1,5} \ ,\ \ \   
\qquad 
H _{24}\equiv H_{15}\ , 
}
where $C_{\rm w} $ is the $\rm w$-component   of the  RR vector potential.
This  background 
 interpolates between the   D2-D4  solution ($\mathrm p=1,\mathrm q=0$)  and  
type IIA   NS1-NS5   solution  ($\mathrm p=0,\mathrm q=1$). 
Its  near-horizon limit is  T-dual to \rf{ee7}--\rf{e9}
\al{\la{e14}
&ds_{10A}^{2}=\td L_s^2\Big[
ds^2_{\ads_3}+ds^2_{S^3}
+\bar{\mathcal Q}_4^{-1}d\y_i^2\Big]
+ \td L_s^{-2} {\bar{\mathcal Q}_4}\,  d\w^2\ ,
\qquad
\td L_s^2=(\cos^2\gamma\,\bar{\mathcal Q}_2+\sin^2 \gamma \,\bar{\mathcal Q}_4)^{1/2}\bar{\mathcal Q}_4^{1/2},
\\
&\mathrm F_4=2\cos\gamma\,g_s^{-1}\, \bar{\mathcal Q}_4\, (i\, \text{vol}_{\ads_3}+\text{vol}_{S^3})\wedge d\w,
\qquad\qquad
\H_3=2\sin\gamma\, \bar{\mathcal Q} _4
\,(i\, \text{vol}_{\ads_3}+\text{vol}_{S^3})\ ,\la{e15} \\ 
&C_{\rm w}=\frac{\cos\gamma \sin\gamma\ (\mathcal Q_4-\mathcal Q_2)}{\cos^2\gamma\ \mathcal Q_2+\sin^2\gamma\ \mathcal Q_4}\ ,
\qquad\ \ \qquad\ \qquad  \  \ \ \quad 
e^{2\phi}=g_s^2\frac{(\cos^2\gamma\ \mathcal Q_2+\sin^2\gamma\ \mathcal Q_4)^{3/2}}{\mathcal Q_2\mathcal Q_4^{1/2}}\ .
\la{e16}}

\newpage
 \subsection{Rotated  M2-M5    background }

Let us now  discuss    how the  11d  uplift    the type IIA   solution \rf{e14}--\rf{e16}  is related 
to  the original   M2-M5 solution \rf{2.1}--\rf{2.3}.
 There  we had    (for   notation see also  \rf{a11}--\rf{a00})
   \be 
   \y_p=(\y_i, \y_4)\ , \quad 
   \y_4\equiv \y_4 + 2 \pi\ls  \vv_4\ , \quad 
   \w\equiv \w+ 2 \pi \ls \uu \ , \quad \vv_4= \ls^{-1} \lpl v_4 = \gs  \ , \quad 
    g'_s = \uu^{-1} \gs =  \uu^{-1} \vv_4 \ , \la{ee160}\ee
   where 
     $\y_4$   played  the role of $x_{11}$   in the reduction to \rf{C.1}--\rf{C.3})  and 
    $g'_s$  is  ``T-dual''  type IIB  coupling.
    Let   us   perform the  following rotation 
     on  these  2-torus 
    coordinates  with radii $\vv_4$ and $\uu$: 
\al{
\y_4 =\cos\gamma\, \tilde{\y}_4 -\sin \gamma\, \tilde{\mathrm w},\qquad \mathrm w=\sin \gamma\, \tilde{\y}_4+\cos \gamma\,\tilde{\mathrm w},\qquad\quad 
 \cos \gamma=\frac{\mathrm p}{\sqrt{\mathrm p^2+\frac{\uu^2}{\vv_4^2}\mathrm q^2}}\ ,  \qquad 
  \pp,\ \qq\in \mathbb Z\ . \label{e29}
}
This  effectively 
  aligns the  new 11th   direction  $\tilde \y_4$  with the  $(\mathrm p,\mathrm q)$-cycle of the   2-torus. It will  then 
  have the period $2\pi \td \vv_4$  where 
  $\td \vv_4 = \sqrt{ \pp^2 \vv_4^2+\qq^2 {\uu^2}}$
   (cf. \rf{B.6}),   so that  
   \be \la{e255}
   R_{11} =\td  \vv_4 \ls , \ \qquad 
 \td  \vv_4 =   \td  g_s  = \gs  \sqrt{\mathrm p^2+\frac{\uu^2}{\vv_4^2}\mathrm q^2}= \sqrt{\pp^2 \vv_4^2 +\qq^2 {\uu^2}}\ , \qquad 
    \lpl =  \td  g_s^{\ftr}   \ls \ . 
   \ee
   Note that $\lpl$ is now related to $\td g_s$  while $\ls$  is  unchanged   (cf. \rf{B.6}). 
The  M2-M5 solution (\ref{2.1})--(\ref{2.4})  written in terms of  the  transformed   coordinates in \rf{e29} 
 takes the following form 
\al{
ds_{11}^{2}=&H_{2}^{-\ftt}H_{5}^{-\ftr}\big(dx^2_{0}+dx^2_{1}\big)
+H_{2}^{\ftr}H_{5}^{\ftt}\big(dr^2+r^2 ds_{S^{3}}^{2}\big)
+H_2^{\ftr}H_5^{-\ftr}d\y_i^2+H_2^{\ftr}H_5^{\ftt}H_{25}^{-1}d\tilde \w^2\nonumber\\
&+H_2^{-\ftt}H_5^{-\ftr}H_{25}\Big[d\tilde \y_4+{\cos \gamma \sin \gamma\, (H_5-H_2)} H_{25}^{-1}d\tilde \w\Big]^2\ , 
\label{e30}\\
F_{4}=&dC_{3}=\big(i\, dH_{2}^{-1}\wedge dx_{0}\wedge dx_{1}-*_{4}dH_{5}\big)\wedge\big(\cos\gamma\, d\tilde \w + \sin \gamma \,d\tilde \y_4 \big)\ ,\label{e31}\\
   H_{2}=&1+\frac{\C_{2}}{r^{2}},\qquad\quad  H_{5}=1+\frac{ \C_{5}}{r^{2}},\quad\qquad  H_{25} \equiv \cos^2\gamma\,H_2+\sin^2 \gamma \,H_5\ .\label{e32}
}
Here $ \C_2$ and $ \C_5$  are  defined  as in the original solution in \rf{24},\rf{244},\rf{2.4},\rf{2.3}
 but   with  $\lpl$   now related to rescaled  type IIA string coupling  as in \rf{e255}  so that 
(cf. \rf{a00})
\al{
 \la{e299}
&  \C_{2}={\lpl^{2}\hv^{-1}\, Q_{2}} = \lpl^{3} \ls^{-1} \vv^{-1}  Q_{2} \to \  \td g_s \ls^2 \vv^{-1} Q_2 = \sqrt{\pp^2 \vv_4^2 +\qq^2 {\uu^2}}\ \ls^2 \vv^{-1} Q_2
 \ , \\
 & \C_{5}={\lpl^{2} u^{-1}\,  Q_{5}}= \lpl^{3} \ls^{-1} \uu^{-1}  Q_{5}\ \to \ 
 \td g_s \ls^2 \uu^{-1} Q_5 = \sqrt{\pp^2 \vv_4^2 +\qq^2 {\uu^2}}\ \ls^2 \uu^{-1} Q_5\ . \la{e300}
}
Since $\gs'= \uu^{-1} {\vv_4}= \uu^{-1} \gs$  these  coefficients 
are thus the same as $\bar \C_1$ and $\bar \C_5$ in \rf{e5} or 
$\bar \C_2$ and $\bar \C_4$ in \rf{e11}.
One can then  check that \rf{e30}--\rf{e32}   reduced to 10d along $x^{11}=\td \y_4$  according to \rf{B.3} 
gives precisely the type IIA   background in \rf{e10}--\rf{e11}. 

The near-horizon limit of \rf{e30}--\rf{e32}  can be equivalently  obtained directly from \rf{2.5},\rf{2.7}
by   performing the rotation \rf{e29}  there.
The  
 result can be written as  
\al{ds_{11}^2=& ( \C_2 \C_5^2)^{\ftr}(ds^2_{\ads_3}+ds^2_{S^3})+(\C_2\C_5^{-1})^{\ftr} d\y_i^2
+ (\C_2\C_5^{-1})^{\ftr} (\cos^2\gamma +\sin^2\gamma\, \C_2^{-1}  \C_5)d\tilde \y_4^2\nonumber\\
&
+  (\C_2\C_5^{-1})^{-\ftt} (\cos^2\gamma   + \sin^2\gamma\, \C_2  \C_5^{-1}  )d\tilde \w^2
+2 (\C^{2}_2\C_5)^{-\ftr}\, \cos\gamma\,\sin\gamma\,(\C_5-\C_2) d\tilde\w d\tilde \y_4\ ,\la{e07} \\
F_{4}=&dC_3=2 \C_5(i\, \text{vol}_{\ads_3}+\text{vol}_{S^3})\wedge(\cos\gamma\, d\tilde \w + 
 \sin \gamma \,d\tilde \y_4 )\ .\la{e08}}
Dimensional reduction along the $x_{11} =\tilde \y_4$ direction then gives the mixed  IIA background in \rf{e14}--\rf{e16}. 

\newpage

 \subsection{AdS$_2\times S^1$ M2 brane   solution and 1-loop partition function }

Let us now consider  a generalisation of the M2 brane solution  discussed  in section 3 to the 
 case of $\gamma\not=0$.  We shall 
 assume that M2  brane 
  is  stretched  along a 2-surface in AdS$_3$ (ending on a circle at the boundary)  and  is wrapping  the 11th dimension 
   $\tilde \y_4$
   so  that  double dimensional reduction will  give a string solution  in the  mixed flux type IIA background \rf{e14}--\rf{e16}.
   
   We shall parametrize AdS$_3$     as in \rf{3.7},\rf{D.3}, i.e. 
   \be \la{e09}
   ds_{\AdS_3}^{2}= \cosh^2 \z \ ds^2_{\AdS_2}  +  d\z^2 \ , \qquad \qquad 
   z=  2 \tanh {\z\ov 2}  \ . 
\ee 
We shall     choose the first   two  of the  M2 brane 
   world volume   coordinates  $\s_1,\s_2$  to be  identified  again  with the AdS$_2$   coordinates and  assume that  
    $\td \y_4\equiv \td \y_4 + 2\pi \td \vv_4\,\ls$  is  proportional to $\s_3\equiv \s_3 + 2\pi $.  We  then set 
   \be  
   \z=\z(\s_1,\s_2) \ , \qquad \qquad  \tilde \y_4=\td \vv_4\, \ls\,   \s_3 \ , \qquad \qquad  \td \w=0,\ \  \y_i=0 \  ,  \la{e100}
\ee
 and also  assume that $S^3$  coordinates are trivial  (i.e.     the brane  is  localized at a point in $S^3$). 
 Equivalently, one may   start  with  the original ``unrotated'' background in \rf{2.5}   and  consider more general M2  brane solution 
 with 
 $\y_4 = \ls\, \td \vv_4\, \cos \gamma \,   \s_3  , \ \  \w= \ls\,  \td \vv_4 \, \sin \gamma \, \s_3 $
  corresponding  to $\tilde \y_4=\td \vv_4\, \ls\,   \s_3, \  \td \w=0$.

  It is   straightforward to check that all the M2-brane equations of motion 
  following from the action in \rf{A.1},\rf{A11} in the background \rf{e07},\rf{e08}, i.e. ${1\ov \sqrt h} \del_m ( h^{mn} \del_n X) \sim F_4 \del X \del X \del X $
  are indeed satisfied by the ansatz \rf{e100}  with the remaining  differential equation for $\z$  also  solved
   by a constant $\z$:
  \al{
&\z= \z_0 \ , \qquad 
\tanh  \z_0  =  -  {   \sin \gamma\,  (\C_2^{-1} \C_5)^{1/2} \,   \ov (\cos^2\gamma   + \sin^2 \gamma \,  \C_2^{-1}  \C_5 )^{1/2}  }\ ,   \la{e013}
 \\
&  \  \cosh^2 \z_0=\X\ , \ \qquad \qquad  \  \X\equiv  1 +  \tan^2 \gamma \,  \C_2^{-1}  \C_5  \ .  \la{e014}
}
  For  $\qq=0$ or $\gamma =0$  (pure RR case)  this reduces to the   solution with $\z=0$   discussed in section 3.
  In the opposite pure NSNS case  $\pp=0$,  $\gamma ={\pi \ov 2}$   the solution  becomes  singular ($\z_0\to \infty$). 
  
  The  corresponding   induced metric is again \AdS$_2\times S^1$    but with rescaled radii  (cf. \rf{55},\rf{2.5},\rf{2.7},\rf{3.5},\rf{3.11}) 
  \al{ \la{e19}
&  (ds^2)_{_{\rm M2}} = \X(  L^2\, ds^2_{\ads_2}   + \tR^2 \td v_4^2\,  d\s_3^2 )= \X L^2\, \big(
ds^2_{\ads_2}   +  \tk^{-2}  d\s_3^2\big)  \ ,   \\ 
&  L^2 =   ( \C_2 \C_5^2)^{\ftr} \, ,\la{e20}\qquad \ \ \   \tR^2 =  \cos^2\gamma\,  \, R^2  \ , \qquad \ \ 
R^2 = \lpl^2 (\C_2\C_5^{-1})^{\ftr}\ , 
\qquad  \ 
\\
&  \tk^{-1} = L^{-1} \tR \, \td v_4 
= \cos \gamma\, \sqrt{ \pp^2  + \qq^2 v_4^{-2}  u^2}\ \k^{-1}  =  \pp\,  \k^{-1}  \,,  \ \ \ \qquad  \k^{-1} =  L^{-1} R\, v_4 
= \gs^{2/3} u^{1/2} Q_5^{-1/2}  \ ,  \la{e22}\\
& 
 \tan^2\gamma = v_4^{-2} u^{2}  \pp^{-2}\qq^{2}  \ ,  \qquad \ \ \ \ 
\td v_4  =   \sqrt{ \pp^2 v_4^2 + \qq^2  u^2} \ . \la{e21}
  }
  Here $ds^2_{\ads_2}(\s_1,\s_2)$ is the unit-radius \AdS$_2$   metric. 
       To compute the  value of the classical  M2 brane action in \rf{A.1},\rf{A11} 
     on the solution \rf{e100},\rf{e013}   we shall assume that the gauge for the field $C_3$  is chosen so that  it vanishes for constant $\z$.\foot{One can   find such gauge    by starting with \rf{e08}  and choosing a specific parametrization   of \AdS$_2$ part of \AdS$_3$ in \rf{e09}. 
       }
       Then the   action is given just  by the  volume part in \rf{A.1} 
  generalising the $\gamma=0$  expression in \rf{3.3}:
  \al{ \la{e23 } 
 \td S_{_{\rm M2}} &= T_2 \, \X^{3/2} L^3 \,   ( 2\pi \tk^{-1}) \,  \vol(\AdS_2) = 
\pp  \, \X^{3/2}  \, S_{_{\rm M2}}\ , \qquad \ \ \ \ \ \ 
  S_{_{\rm M2}} =-   { v_4\ov \sqrt{\hv u} } \sqrt{{Q_2 Q_5}} \ .
}
Let us now  consider the type IIA string theory expressions for
  the  corresponding  AdS$_2$   string solution in  the  IIA background in \rf{e14}--\rf{e16}. 
  From  \rf{e14}--\rf{e16}  we  get for the induced string metric and  the  string action  value on  the solution 
         (here $Q_4=Q_5$)\foot{The same  value for $\z_0$ in \rf{e014}   is of course  found also by solving 
           the   string equations directly  in  the  background \rf{e14},\rf{e15}. }
\al{
&(ds^2)_{s} =   \hat L_s^2 ds^2_{\ads_2} \ , \ \ \ \  \ \ \ \ \   \hat L_s^2 = \X \td  L^2_s= 
\pp\,   \X^{3/2} 
   (\C_2\C_4)^{1/2} \ , 
\la{e160}\\
&\C_2= \ls^2\gs    \vv^{-1} Q_2,\ \ \  \ \ \C_4= \ls^2 \gs \uu^{-1} Q_4\ , \ \ \ \ \ 
 \la{e66}\qquad   
{\td {\rm g}_s}^2 = e^{2\phi}=g_s^2\cos^3 \gamma\,   \X^{3/2}  (\C_2^{-1}  \C_4)^{-1/2} \ ,\\  
& 
\td S_s=- 2 \pi  T \hat L_s^2 = -  \ls^{-2}  \hat L_s^2 
 = \pp\,  \X^{3/2}   S_s \ , \qquad  \ \ \ 
 S_s=-   \ls^{-2}   L_s^2  =- \gs 
(  \vv^{-1} \uu^{-1} Q_2  Q_4)^{1/2} \ .   \la{e111}}
The value of the string action  is the same  as the M2  brane one  in \rf{e21} (cf. \rf{155}). 

 In the special case   of  $\Q_2=\Q_4$
 the corresponding  string solution   can be  shown to  be  equivalent to 
 the one  
  discussed in  ref.  \ci{Pajer:2021bfr}   which  considered the    string  representing an analog of WL 
   in the  type II AdS$_3\times S^3 $    background   supported by the 
 $\H_3=2 \hq  \td L^2_s \, \vol_{\ads_3}$  flux   with $\td L_s$ assumed  to be independent of $\hq$.
  The world  surface   ended  on a circle at the boundary of \AdS$_3$  and had AdS$_2$  induced metric.
 When  $\C_2=\C_4\equiv \C$  the  type IIA dilaton  in  \rf{e16} is  $\gamma$-independent,  
  $e^{2\phi}=  g^2_s $, and 
     using \rf{e29},\rf{e014},\rf{e160} we get\foot{
Equivalent expressions are found also  by starting with the type IIB   background in \rf{ee7}--\rf{e001}.}
 \al{ \X= (\cos \gamma)^{-2} \ , \ \ \ \ \ \ 
 \hat L_s^2 =  \X \tL_s^2 =\pp (\cos \gamma)^{-3} \, L_s^2  \ , 
 \la{e210} \qquad \ \   \td L^2_s = \pp (\cos \gamma)^{-1} \, L_s^2
  \ , \qquad \ \ \   L_s^2 =  \C\ .
  }
The NSNS flux $\H_3$   in \rf{e15} is  given by 
\be \la{e48}  \H_3=2 h_3
\,(i\, \text{vol}_{\ads_3}+\text{vol}_{S^3})  \ , \qquad \qquad 
h_3 = 
 \sin \gamma\, \bar \C_4= 
 \hq\, \td L^2_s  \ , \qquad\qquad   \hq \equiv \sin \gamma\,  
  \ .    
\ee
The string action in \rf{e111}   written in terms of $\hq$ and the  radius   $\td L_s$   in \rf{e14}  takes the same form as in \ci{Pajer:2021bfr}\foot{Note that in  the discussions of the 
 mixed flux case in  \ci{Pajer:2021bfr} and, e.g.,  \ci{Hoare:2013pma,Hoare:2022asa} 
 the radius $\td L_s$ of AdS$_3 \times S^3$  part of type IIA metric in \rf{e14}   was treated as an independent  parameter,  
 with the NSNS and RR   fluxes proportional to $ \hq\, \tL_s^2$ and $ \sqrt{1-\hq^2}\, \tL_s^2$  respectively. 
 This is different from the parametrization used in  \rf{e002}--\rf{e001}  where  both $Q_2$ and $Q_4$ 
 were  assumed to be quantized with $\td L$  expressed in terms of them.}  
\be \la{e44} \td S_s 
  = \pp (\cos \gamma)^{-3} \,  S_s  (L_s) 
= {1\ov 1-\hq^2}\,   S_c (\td L_s) \ , \ \ \ \ \ \ \ \ \ \   S_c (L_s) \equiv  -   \ls^{-2}   L_s^2    \ . \ee
  The   corresponding 1-loop correction to the  string partition function  turns out 
  to have  the same form  as in
 the pure RR case  \rf{66}   but with the effective string tension  rescaled like the action in \rf{e44}  by $ (1-\hq^2)^{-1}$   \ci{Pajer:2021bfr}.

 Starting with M2  action in the rotated  background \rf{e07},\rf{e08}  it is  
 straightforward  to  expand it near the  M2  brane solution \rf{e100},\rf{e013} to quadratic order in fluctuations
  using the same static    gauge  as in section 3 (i.e. setting  the fluctuations of AdS$_2$ coordinates and of $\td \y_4$ to zero).  
  After a rescaling of the fluctuation fields  we find  that the bosonic  Lagrangian  depends just on one parameter -- 
  it  is the same as in \rf{3.16}  with 
  $\k$ replaced by $\tk= \pp^{-1} \k$  defined  in \rf{e22}.\foot{
  This  may not be too surprising given that  the background   \rf{e07}, \rf{e08} 
  is  just a coordinate transformation of the original one and the  M2 brane solution still having AdS$_2 \times S^1$  induced metric.}
  As a result,  the bosonic  contribution 
  to the 1-loop M2  brane   partition function is given  again by \rf{3.20},\rf{318} where 
  all the masses  depend  only on  $\tk$.   Similarly, the fermionic   contribution  is found to be 
  given by \rf{3.32} 
  with  $\k \to \tk$  so that the total  is   again \rf{4.42} with 
  \be \la{e77}
   \td M^2_{n,\pm}
    =1 + \tk^2 n^2 \pm  \sqrt{1+4 \tk^2 n^2}\ , \qquad \qquad  
    \td  \M_{n,\pm }
    =  \tfrac 12 \big(\sqrt{1 +   4 \tk^2 n^2} \pm 1  \big)   \ .   \ee    
 As a result,  we find the same   expressions as in  section 4. In particular,  for $\tk^2 > {3\ov 4}$ we get  
 as in \rf{3.64} 
 \begin{equation}
    \hat Z_1=\frac {\tk}{\sqrt{2\pi}}=  \frac {\k}{\pp\sqrt{2\pi}} \ .\label{3645}
\end{equation}
 Thus the 1-loop   result  is the same as in the 10d string theory limit  (cf. \ci{Pajer:2021bfr}) and 
 does not depend  on  values of $\qq$  and $Q_1$, 
 with  the pure NSNS limit of $\pp=0$   being   singular. 
 Note  that 
instead of $\pp,\qq,Q_2,Q_4$  both  the  classical action \rf{e111} and the 1-loop correction \rf{3645} 
can be expressed in terms of the 4  integer  charges  in \rf{e567}  (with $Q_2=Q_1, \, Q_4=Q_5$). 

\newcommand{\Hf}[1]{H^{{#1}}_5}
\newcommand{\Hfp}[1]{H'^{{#1}}_5}
\newcommand{\Ht}[1]{H_2^{#1}}

\section{$\ads_3\times S^3\times S^3 \times S^1\times S^1$ case}

  In $D=10$  the    NS-NS  $\ads_3\times S^3\times S^3 \times S^1$ background   
    has  straightforward   description  in terms of WZW model  and is  also 
    a near-horizon limit of F1-NS5-NS5  configuration that has an 11d uplift to M2-M5-M5'
      with  $F_4$  given  by  NS-NS 3-form times  $dx_{11}^2$  (see, e.g.,   \ci{Boonstra:1997dy,Gukov:2004ym}). 
       Interchanging the roles  of the two $S^1$'s  one gets  an 11d    solution that reduces to  10d   background 
    supported by  R-R  $F_4$  flux. 
 Below  we extend the  above analysis  of the 1-loop   partition function of  M2 brane on 
   $\ads_2 \times S^1$  to the corresponding    $\ads_3\times S^3\times S^3 \times S^1\times S^1$   background.

 It  can be represented   is a   near horizon limit of  the  following 
 M2-M5-M5 intersection \ci{Boonstra:1997dy}  
 \al{
ds^2=&\Hf{-1/3}\Hfp{-1/3}\Ht{-2/3}(dx_1^2+dx_2^2)+\Hf{2/3}\Hfp{-1/3}\Ht{1/3}(dr^2+r^2 dS_3)\nonumber \\
&+\Hf{-1/3}\Hfp{2/3}\Ht{1/3}(dr'^2+r'^2 dS'_3)+\Hf{2/3}\Hfp{2/3}\Ht{-2/3}d\w^2,\\
F_4=&\left(i d\Ht{-1}\wedge dx_1\wedge dx_2-*_4d\Hf{} -*'_4d\Hfp{} \right)\wedge d\w,\\
\Ht{}=&1+\frac {\Q_2}{r^2r'^2},\qquad \Hf{}=1+\frac {\Q_5}{r^2},\qquad \Hfp{}=1+\frac {\Q'_5}{r'^2}.
}
 $\Q_2,\Q_5,\Q_5'$ are determined by charge quantization to be 
\be \la{ppp}
\Q_2={8Q_2\over Q_5^2Q_5'^2}  \, \lpl^4u^2 v^{-3},\, \qquad 
\Q_5= {Q_5 \lpl^2} u^{-1},\,\qquad \Q_5'= {Q_5' \lpl^2}u^{-1},\qquad \,Q_5, Q'_5 \in\mathbb Z\ .
\ee
In the near horizon limit,
 \al{
ds^2=&L^2ds_{\ads_3}^2 +\Q_5dS_3+\Q_5'dS'_3+d\w^2+R^2d\w'^2,\quad \label{f4}\\
F_4=&2(iL^2 \text{vol}_{\ads_3} +\Q_5\text{vol}_{S^3}+\Q_5'\text{vol}_{S'^3})\wedge d\w,\qquad\qquad  \frac 1{L^2}=\frac{1}{\Q_5}+\frac{1}{\Q_5'},\label{f5}\\
\w\equiv& \w+2\pi \lpl u,\qquad \w'\equiv \w'+2\pi \lpl v,\qquad R^2=L^2\Q_5^{-1}\Q_5'^{-1}(2\pi \lpl)^2 \ , 
}
where $\w' = (2\pi \lpl)^{-1} ( \Q_5 \log r - \Q'_5 \log r')$. 
For (\ref{f4}),(\ref{f5}), we have made the choice $\Q_2=\Q_5\Q'_5$, which relates the two dimensionless parameters $v=2u^2$.
The  supersymmetric   $\ads_2 \times S^1$  M2  brane solution   corresponds to 
$
x_1=\sigma_0,\,x_2=\sigma_1,\, \w'=\lpl v\,\sigma_3=2\lpl u^2\,\sigma_3.
$
The  value of the classical M2  brane action is then
\al{ 
&S_{cl}=T_2L^2R\,(2\pi\lpl v)\, \text{vol}(\ads_2)=\mathrm T_2(4\pi u^2)\, \text{vol}(\ads_2)\ .
}
Equivalently,   
 we may rewrite \rf{f4},\rf{f5}  in terms of the    redefined  coordinates and parameters  as 
\al{
ds^2=&L^2\left[ ds^2_{\ads_3}+d\tilde S_3+d\tilde S'_3+d\tilde \w^2+\hat \kappa^{-2}d \tilde \varphi^2\right] ,\quad\qquad \ \ \ \hat\kappa^{-1}\equiv 2RL^{-1}\, \lpl u^2 ,\\
F_4=&2L^3(i \text{vol}_{\ads_3} + \cos\nu
 \,\text{vol}_{S^3}+ \sin \nu  \,\text{vol}_{S'^3})\wedge d\tilde \w ,\qquad\ \ \ \ \  \cos\nu \equiv \sqrt{\tfrac{\Q_5'}{\Q_5+\Q_5'}}=  
\sqrt{\tfrac{Q_5'}{Q_5+Q_5'}}
\ ,  \\
d\tilde S_3=&\Q_5\,L^{-2}dS_3,\qquad d\tilde S'_3=\Q_5'\,L^{-2}dS'_3,\qquad \tilde \w =L^{-1}d\w,\qquad \tilde \varphi=RL^{-1}\hat\kappa\,\w',\qquad
 \tilde \varphi \equiv \tilde \varphi+2\pi\ . }
 Choosing the static gauge  the 
  bosonic quadratic fluctuation action is the
   same as for the $\ads_3\times S^3 \times T^4 \times S^1$ case. 
   The diagonalised masses are given by (\ref{318}) with $\kappa\rightarrow \hat \kappa$,
\be
\hat M^2_{n,0}=n^2\hat \kappa^2,\qquad \hat M^2_{n,\pm}=
1+\hat\kappa^2 n^2 \pm \sqrt{1+4 \hat\kappa^2 n^2}\ .
\ee
The quadratic  fermionic fluctuation action takes the same form as (\ref{3.27}) with the following replacement
\al{
\tfrac 1{2} \bar \theta\Gamma^{\underline {012\,10}} (1-i\Gamma^{\underline{012345}})
\rightarrow
\tfrac{1}{2} \bar{\theta} \Gamma^{\underline{012\,10}}(1-i \cos\nu \,\Gamma^{\underline{012345}}-i \sin\nu  \,\Gamma^{\underline{012678}}) ,
}
which  corresponds to a  rotation in the two $S^3$'s. This replacement of gamma matrices does not change the fermionic determinant.
As a result, the fermion masses are again given by  (\ref{3.33}) with   $\kappa\rightarrow \hat \kappa$
\be
\mathcal{M}_{n,\pm}=\tfrac{1}{2}\left(\sqrt{1+4 \hat\kappa^2 n^2}\pm1\right).
\ee
The one-loop   partition function is then given  as in  (\ref{4.42}),(\ref{3.47}),(\ref{3.64})  by 
\be
\hat{Z}_1=e^{-\hat{\Gamma}_1}=\frac{\hat \kappa}{\sqrt{2 \pi}},\qquad \qquad \ \ \ \hat\kappa=
\frac{\sqrt{\Q_5 \Q'_5}}{4\pi\lpl^2 u^2}=
\frac{\sqrt{Q_5Q_5'}}{4\pi u^3}.
\ee

\small
\bibliographystyle{JHEP-v2.9}
\bibliography{bib}
\end{document}